\documentclass[12pt]{iopart}

\usepackage{iopams}
\usepackage{epsfig}
\usepackage{graphicx, subfigure}
\usepackage{multirow}
\usepackage{hyperref}
\usepackage{cite}
\usepackage{color}

\begin{document}

\title[Hydogen Atom in Strong Elliptical Laser Field]{Hydrogen Atom in Strong Elliptically Polarized Laser Fields within Discrete-Variable Representation}

\author{Sara Shadmehri$^{1}$ and Vladimir S. Melezhik$^{1,2}$}
\address{$^1$ Bogoliubov Laboratory of Theoretical Physics, Joint Institute for Nuclear Research, Dubna, Moscow Region 141980, Russian Federation}
\address{$^2$ Dubna State University, 19 Universitetskaya St., Moscow Region 141982, Russian Federation}

\eads{\mailto{shadmehri@theor.jinr.ru}, \mailto{melezhik@theor.jinr.ru}}

\vspace{10pt}
\begin{indented}
\item[]9 Dec 2022
\end{indented}



\begin{abstract}
The nondirect product discrete variable representation (npDVR) is developed for the time-dependent Schr\"odinger equation with nonseparable angular variables and is applied to a hydrogen atom in elliptically polarized strong laser fields. The 2D npDVR is constructed on spherical harmonics orthogonalized on the 2D angular grids of the Popov and Lebedev 2D cubatures for the unit sphere. With this approach we have investigated the dynamics of a hydrogen atom initially in its ground state in  elliptically polarized laser fields with the intensity up to $I=10^{14}$ W/{cm}$^2$ and wavelength of $\lambda=800$ nm. For these parameters of the laser field and the entire range of ellipticity variation, we have calculated the total excitation and ionization yields of the atom. The performed analysis of the method convergence shows that the achieved accuracy of our calculations significantly exceeds the accuracy of recent works of other authors relevant to the problem, due to the high efficiency of the 2D npDVR in approximating the angular part of the 3D time-dependent Schr\"odinger equation.
We also propose a new simple procedure for infinite summation of the transition probabilities to the bound states of the hydrogen atom in calculating the total excitation yield and prove its accuracy by comparison with conventional methods. The obtained results show the potential prospects of the 2D npDVR for investigating atomic dynamics in stronger laser fields.
\end{abstract}


\section{Introduction}
Over the past decades, great progress has been made in the development of high-power laser technology, and many laser driven nonlinear processes have been discovered and investigated such as high harmonic generation (HHG) \cite{HHG1,HHG2}, above threshold ionization (ATI) \cite{ATI1,ATI2}, laser-induced tunneling \cite{tunneling1,tunneling2} etc (see, for example, \cite{rev1,rev2,rev3} and references therein).
An adequate quantitative description of all these processes required the development of special methods that take into account the features of quantum dynamics of an electron in a combined Coulomb and time-dependent laser field.
Thus, the electron motion should be described in a very extended space and during long time-scales\cite{3stepM}. The polarization of the laser field plays here the role of an additional control parameter, but it also complicates the problem. While the problem of a hydrogen atom in a linearly polarized laser field is two-dimensional due to cylindrical symmetry and a large set of efficient methods has been developed to solve the corresponding 2D Schr\"odinger equation, in an elliptically polarized field the problem becomes three-dimensional and requires special consideration.

Different aspects of strong laser-atom interactions induced by elliptically polarized laser fields were investigated through the semi-analytical models \cite{Moller2012,Bandrauk2012,Bandrauk2014,Zhao2019} and numerical simulations of the corresponding 3D time-dependent Schr\"odinger equation (3D TDSE) \cite{He2007,Tong2017,Gao2019,Murakami2016}.
Thus, the HHG of a hydrogen atom in elliptically polarized laser fields was studied numerically in \cite{He2007} by integrating the 3D TDSE in Cartesian coordinates while assuming a soft-core Coulomb potential to avoid the singularity at the origin. The effect of the electron-core Coulomb interaction on the dynamics of a hydrogen atom in elliptical laser fields was surveyed in \cite{Gao2019} by using a 3D time propagator pseudospectral method \cite{Tong2017} in spherical coordinates $(r,\theta,\phi)$ with up to $256$ partial waves, where the number of simulation operations was greatly reduced from $N^2$ to $N(N_r+N_\theta+N_\phi)$ ($N=N_r N_\theta N_\phi$ is the total number of nodes in the spatial grid).

In \cite{Melezhik97}, an alternative approach was applied to calculate the HHG by a hydrogen atom in an elliptically polarized laser field,  using the discrete variable representation (DVR) to approximate the angular part of the 3D TDSE, which demonstrated fairly fast convergence and had a sufficiently high computational efficiency with the number of operations $\sim N_r (N_\theta N_\phi)^\gamma$, where $1\lesssim \gamma < 2$ \cite{Melezhik2002}. Note a well-known advantage of the DVR \cite{DVR1,DVR2,DVR3}: any local spatial interaction is diagonal in DVR, being approximated by its values at the grid points, and the only off-diagonal elements belong to the kinetic energy operator. In \cite{Melezhik97}, the DVR was constructed on spherical functions orthogonalized on the 2D angular grids constructed as a direct product of the nodes of 1D Gaussian quadratures over the angular variables $\theta$ and $\phi$. In our recent work \cite{Shadmehri2020}, it was shown that 2D nondirect product DVR (2D npDVR) essentially improves approximation of the nonseparable angular part of the 3D stationary Schr\"odinger equation and accelerates the method convergence by using as an example a hydrogen atom in crossed electric and magnetic stationary fields. For constructing the 2D npDVR we have used the Popov \cite{Popov1,Popov2,Popov3} and Lebedev \cite{Lebedev1,Lebedev2,Lebedev3} cubatures for the unit sphere which are invariant under the icosahedral and octahedral rotation groups, respectively. They belong to the class of most efficient quadrature rules (cubatures) for the unit sphere invariant under finite rotation
groups outlined by Sobolev in his seminal works \cite{Sobolev1,Sobolev2}.

In the present work, our 2D npDVR proposed in \cite{Shadmehri2020} is extended to the 3D TDSE with the strongly nonseparable angular part and its capabilities are demonstrated for describing the dynamics of a hydrogen atom in an elliptically polarized strong laser field. While, once we investigated HHG for hydrogen atoms in elliptical laser fields through the Gaussian scheme \cite{Melezhik97}, here, using both DVR based on the 1D Gaussian angular grids  and 2D npDVR , we focus on the ionization and excitation processes which are fundamental parts of many photoabsorption phenomena \cite{Photoabsorp1,Photoabsorp2,Photoabsorp3}. More rapid convergence of the 2D Popov and Lebedev  npDVRs than the 2D Gaussian DVR is demonstrated in these problems. In contrast to Ref.\cite{Tong2017}, where the computational time has a quadratic dependence on $N_r$, in our 2D npDVRs  it is linear \cite{Melezhik2002} and proportional to $ N_r(2 N_\Omega^2+\alpha N_\Omega)$  where  $N_\Omega$ denotes the total number of angular grid points, which is the product $N_\Omega=N_\theta\times N_\phi$ for the 2D Gaussian DVR. This allowed us to significantly improve the accuracy of calculations of the probabilities of excitation and ionization of a hydrogen atom in strong elliptically polarized laser fields, achieved in the best recent calculations of other authors\cite{Tong2017,Gao2019}. We also suggest a novel approach to infinite summation over the bound states of a hydrogen atom in calculating the total excitation probability and prove its accuracy by comparing with conventional methods in the case of linearly polarized laser fields. The demonstrated efficiency of our method in these problems makes it promising in stronger fields for solving the corresponding 3D TDSE with the nonseparable angular part.

In the next section, our approaches to numerical integration of the 3D TDSE with the 2D Gaussian DVR and 2D Popov and Lebedev npDVRs are presented. We also describe the formulas for calculating the ionization yield and the procedure for approximating infinite summation over the states of the discrete spectrum of the hydrogen atom when calculating the total excitation probability. In Sec. \ref{Sec:Results}, the results of calculations and discussions are presented. The concluding remarks are given in the last section.
\section{Theoretical Method}\label{Sec:Theory}
\subsection{Time-dependent Schr\"odinger Equation in 2D DVR}\label{Subsec:TDSE-theory}
To study the dynamics of a hydrogen atom in an arbitrarily polarized laser field, we need to solve the corresponding TDSE (atomic units $m=e=\hbar=1$ are used throughout the text unless stated otherwise),
\begin{eqnarray}\label{eq:mainTDSE}
i\frac{\partial}{\partial t}\psi({\bf r},t)=\left[H_{0}+V({\bf r},t)\right]\psi({\bf r},t)
\end{eqnarray}
with the Hamiltonian $H_{0}=-\frac{1}{2}\Delta-\frac{1}{r}$ of the hydrogen atom in free space. We describe the interaction of the atomic electron with the laser field
\begin{eqnarray}
{\bf E}(t)=\frac{f(t)}{\sqrt{1+\epsilon ^2}}E_0\left(\hat{\bf x}\cos\omega t+\hat{\bf y}\epsilon\sin\omega t\right)
\end{eqnarray}
in the length gauge under the dipole approximation
\begin{eqnarray}
V({\bf r},t)=-{\bf r}\cdot{\bf E}(t) \,\,,
\end{eqnarray}
where $E_0$, $0 \leq \epsilon \leq 1$ and $\omega$ are the electric field amplitude, ellipticity and frequency of the laser field, respectively. The pulse envelope function is considered as
\begin{eqnarray}
f(t)=\cos ^2(\frac{\pi t}{n_TT})~~~,~~~-n_TT/2<t<n_TT/2\,\,,
\end{eqnarray}
 where the $n_T$ optical cycles of the period $T=2\pi/\omega$ are included during the pulse.

To solve TDSE, we follow the  2D DVR method \cite{Melezhik97,Melezhik2002,Melezhik99,Melezhik2016} where the electron wavefunction $\psi({\bf r},t)$ in the spherical coordinates $(r,\Omega)=(r,\theta,\phi)$ is expanded as
\begin{eqnarray}\label{eq:main-psi-expansion}
\psi(r,\Omega, t)=\frac{1}{r}\sum_{j=1}^{N_{\Omega}}f_j(\Omega)u_j(r,t)
\end{eqnarray}
over the basis $f_j(\Omega)$ associated with a 2D angular grid $\Omega_j$ ($j=1,2,..., N_{\Omega}$) on the unit sphere $\Omega$. In the works of V.S. Melezhik on 2D DVR in application to TDSE, the angular grids $\Omega_j=(\theta_{j_{\theta}},\phi_{j_{\phi}})$ were constructed as direct products of the nodes $\theta_{j_{\theta}}$ and $\phi_{j_{\phi}}$ of the 1D Gaussian quadratures over the $\theta$ and $\phi$ variables, respectively, where $j_{\theta} = 1,...,N_{\theta}$, $j_{\phi}=1,...,N_{\phi}$ and $N_{\Omega}=N_{\theta}\times N_{\phi}$ \cite{Melezhik97,Melezhik2002,Melezhik99}. This construction turned out to be very efficient in integration of the TDSE with three \cite{Melezhik97,Melezhik99,Melezhik2000} and four \cite{Kim,Melezhik2007,Melezhik2009} spatial variables. Particularly, with this approach the polarization of the high harmonics generated by hydrogen atoms were successfully investigated in elliptically polarized laser fields \cite{Melezhik97}. Recently \cite{Shadmehri2020}, we have improved the computational scheme for the 3D stationary Schr\"odinger equation by implementing the 2D Popov and Lebedev nondirect product DVRs on the unit sphere $\Omega_j =(\theta_j,\phi_j)$ instead of the Gaussian 2D DVR. When integrating the stationary 3D Schr\"odinger equation for the hydrogen atom in crossed magnetic and electric fields, it was shown that the use of the 2D npDVR essentially accelerated the convergence of the computational scheme over $N_{\Omega}$ with respect to the 2D DVR scheme based on the Gaussian 1D quadratures with $N_{\Omega}=N_{\theta}\times N_{\phi}$ . This result motivated us to extend the 2D npDVR based on the Popov and Lebedev cubatures to the TDSE (\ref{eq:mainTDSE}) for the hydrogen atom in a strong elliptically polarized laser field where convergence of the computational scheme becomes crucial due to complexity of the problem to be solved.

When angular grid points $\Omega_j=(\theta_{j_\theta},\phi_{j_\phi})$ are chosen as nodes of 1D Gaussian quadratures over the $\theta$ and $\phi$ variables, the basis functions $f_j(\Omega)$ are defined as
\begin{eqnarray}\label{eq:fj_Gaussian}
f_j(\Omega)=\sum_{\nu=1}^{N_{\Omega}}\bar{Y}_\nu(\Omega)[Y^{-1}]_{\nu j}\,.
\end{eqnarray}
Here the index $\nu$ numerates the pair $\{l,m\}$ ($m=-(N_\phi-1)/2,..,+(N_\phi-1)/2$ and $l=|m|,..,|m|+N_\theta-1$) related to the modified spherical harmonic
\begin{eqnarray}\label{eq:Ybar_nu}
\bar{Y}_\nu(\Omega)=\bar{Y}_{lm}(\Omega)=e^{im\phi}\sum_{l^\prime}{c_{l}^{l^\prime}}P_{l^\prime}^{m}(\theta)\,,
\end{eqnarray}
where $P_{l^\prime}^{m}(\theta)$ are the associated Legendre polynomials. For the majority of $l$ and $l^\prime$ from a given above set, except the largest $l$ and $l^\prime$, $c_{l}^{l^\prime}=\delta_{ll^\prime}$ and $\bar{Y}_{lm}(\Omega)$ coincide with the spherical harmonics. The non zero $c_{l}^{l^\prime}$ for the largest $l$ and $l^\prime$ are found using the Gram-Schmidt orthogonalization procedure for $\bar{Y}_{lm}(\Omega)$ on the grid $\Omega_j$ \cite{Shadmehri2016}. The matrix $Y^{-1}$ is inverse to the $N_\Omega\times N_\Omega$ matrix $Y_{j\nu}=\sqrt{w_j}\bar{Y}_{\nu}(\Omega_j)$, where $w_j=(2\pi{w_j}^\prime)/N_\phi$ and ${w_j}^\prime$ are the weights of the Gaussian quadrature over the $\theta$ variable.

On the other hand, if the Popov or Lebedev cubature is used for generating the angular grid points $\Omega_j=(\theta_j,\phi_j)$ with $j=1,..,N_{\Omega}$, then $f_j(\Omega)$ is defined as
\begin{eqnarray}\label{eq:fj_Popov}
f_j(\Omega)=\sum_{\nu=1}^{N_\Omega}\sqrt{4\pi w_j}\Phi_\nu(\Omega)\Phi^{*}_{\nu}(\Omega_j)\,\,,
\end{eqnarray}
where $w_j$ is the Popov or Lebedev weight ($\sum_{j=1}^{N_\Omega}{w_j}=1.0$) (for Lebedev and Popov grid points and weights see \cite{Burkardt} and \cite{Popov1,Popov2,Popov3} , respectively). The function $\Phi_\nu$ is a special linear combination of the spherical harmonics, $\Phi_\nu(\Omega)=\sum_{\mu}S_{\nu\mu}Y_\mu(\Omega)$ obtained within the method described in our previous work \cite{Shadmehri2020}. Here $\hat{S}$ refers to a $N_\Omega\times \bar{N}_\Omega$ matrix ( with $\mu=1,...,\bar{N}_\Omega \geq N_\Omega$ ) which makes the set of $N_\Omega$ basis functions $\Phi_\nu(\Omega)$ orthonormal over the grid points $\Omega_j$ in the sense of the Popov or Lebedev cubatures.

Thus, the problem is reduced to a system of Schr\"odinger-type equations which should be solved for the unknown $N_\Omega$-dimensional vector ${\bf u}(r,t)=\{\sqrt{w_j}u_j(r,t)\}_1^{N_\Omega}=\{\sqrt{w_j}r\psi(r,\Omega_j,t)\}_1^{N_\Omega}$,
\begin{eqnarray}\label{eq:Sch-type_Eq}
i\frac{\partial}{\partial t}{\bf u}(r,t)=\left[\hat{H}_0(r)+\hat{V}(r,t)\right]{\bf u}(r,t)\,\,,
\end{eqnarray}
where the elements of the $N_\Omega\times N_\Omega$ matrices $\hat{H}_0(r)$ and $\hat{V}(r,t)$  are represented as
\begin{eqnarray}
H_{0jj^\prime}(r)&=&-\frac{1}{2}(\delta_{jj^\prime}\frac{d^2}{dr^2}-\frac{L_{jj^{\prime}}^2}{r^2})-\frac{1}{r}\delta_{jj^\prime}\,,\label{eq:H0_element}\\
V_{jj^\prime}(r,t)&=&V(r,\Omega_j ,t)\delta_{jj^\prime}\,.\label{eq:V_element}
\end{eqnarray}
In the Gaussian case, the elements of the ${\hat{L}}^2$ matrix are defined as
\begin{eqnarray}\label{eq:L2-gaus}
L_{jj^{\prime}}^2=\sum_{\nu=1}^{N_\Omega}{\sqrt{w_jw_{j^\prime}}}\bar{Y}_\nu(\Omega_j)l(l+1){\bar{Y}_\nu}^\ast(\Omega_{j^\prime})\,.
\end{eqnarray}
whereas in the Popov and Lebedev cases, they are written as
\begin{eqnarray}\label{eq:L2-popov}
L_{jj^\prime}^2=4\pi\sum_{\nu=1}^{N_\Omega}{\sqrt{w_jw_{j^\prime}}}\sum_{\mu=1}^{\bar{N}_\Omega}S_{\nu\mu}l_\mu(l_\mu+1){{Y}_\mu}(\Omega_{j}){\Phi}_\nu^\ast(\Omega_{j^\prime})\,,
\end{eqnarray}
with $\bar{N}_\Omega \geq N_\Omega$ and $\mu=\{l_\mu,m_\mu\}$ defined in \cite{Shadmehri2020}.

To propagate the wavefunction in time $t_n\rightarrow t_{n+1}=t_n+\Delta t $, the component-by-component split-operator method \cite{Marchuk} is applied as
\begin{eqnarray}\label{eq:splitted-Eq}
\fl{\bf u}(r,t_n+\Delta t)=\exp [-\frac{i}{2}\Delta t \hat{V}(r,t_n)]\exp [-i\Delta t \hat{H}_0(r) ]\exp [-\frac{i}{2}\Delta t \hat{V}(r,t_n)]{\bf u}(r,t_n)\,\,.
\end{eqnarray}
Since the interaction potential $\hat{V}(r,t)$ is diagonal in our 2D DVRs over both angular grid distributions (either Gaussian $\Omega_j=(\theta_{j_\theta},\phi_{j_\phi})$ or Popov (Lebedev) $\Omega_j=(\theta_{j},\phi_{j})$), the first and last steps of the procedure(\ref{eq:splitted-Eq}) represent simple multiplications of the vectors $\exp\left[-(i/2)\Delta t V(r,\Omega_j,t)\right]$ and ${\bf u}(r,t)$ which results in ${\bf u}(r,t+\frac{1}{4}\Delta t)$. The intermediate step is performed after diagonalizing the nondiagonal part (${\hat{L}}^2$ operator) which  transforms $\hat{H}_0$ to a diagonal matrix $\hat{A}$. Then we can approximate its related exponential expression according to
\begin{eqnarray}
\exp [-i\Delta t\hat{A}]\approx {\left(1+\frac{i}{2}\Delta t\hat{A} \right)}^{-1}\left(1-\frac{i}{2}\Delta t\hat{A} \right)
\end{eqnarray}
which ensures the desired accuracy of the numerical algorithm (\ref{eq:splitted-Eq}).

Thus, we reach the following initial value problem for the intermediate step:
\begin{eqnarray}
\left(1+\frac{i}{2}\Delta t\hat{A} \right)\bar{\bf u}(t_n+\frac{3}{4}\Delta t)=\left(1-\frac{i}{2}\Delta t\hat{A} \right)\bar{\bf u}(t_n+\frac{1}{4}\Delta t)\,,
\end{eqnarray}
\begin{eqnarray}
\bar{\bf u}(r=0,t_n+\frac{3}{4}\Delta t)=\bar{\bf u}(r=r_m,t_n+\frac{3}{4}\Delta t)=0~~,~~r_m\rightarrow \infty\,\,.
\end{eqnarray}
Here the transformed wavefunction $\bar{\bf u}$ and operator $\hat{A}$ are defined based on the transformation method which is used for diagonalizing the ${\hat{L}}^2$ operator. When using Gaussian grid points, we apply a simple unitary transformation $M_{j\nu}=\sqrt{w_j}\bar{Y}_\nu(\Omega_j)$ \cite{Melezhik2002,Melezhik99}, then
\begin{eqnarray}
\bar{\bf u} = \hat{M}{\bf u}\,,
\end{eqnarray}
\begin{eqnarray}
\label{eq:matrixA}
\hat{A}_{\nu\nu^\prime}=\left(\hat{M}\hat{H}_0\hat{M^\dagger}\right)_{\nu\nu^\prime}=\left[-\frac{1}{2}\frac{d^2}{ dr^2}+\frac{l(l+1)}{2r^2}-\frac{1}{r}\right]\delta_{\nu\nu^\prime}
\end{eqnarray}
with $\nu=\{l,m\}$.

In the Popov and Lebedev cases, we use the conventional numerical diagonalization method. We find the eigenvectors and eigenvalues of the ${\hat{L}}^2$ matrix in these representations. Then the transformation matrix $\hat{W}$ is constructed from the numerically calculated eigenvectors which are stored on its columns so that $\hat{W}^{-1}{\hat{L}}^2\hat{W}=\hat{J}$, where $\hat{J}$ is the diagonal matrix of the acquired eigenvalues $J_\nu$. Thus $\bar{\bf u}$ and $\hat{A}$ are represented as
\begin{eqnarray}
\bar{\bf u}=\hat{W}^{-1}{\bf u}\,,
\end{eqnarray}
\begin{eqnarray}
\label{eq:matrixA2}
\hat{A}_{\nu\nu^\prime}=\left(\hat{W}^{-1}\hat{H}_0\hat{W}\right)_{\nu\nu^\prime}=\left[-\frac{1}{2}\frac{d^2}{
dr^2}+\frac{J_{\nu}}{2r^2}-\frac{1}{r}\right]\delta_{\nu\nu^\prime}
\end{eqnarray}
with $\nu=1,2,...,N_\Omega$.

Finally, before applying the last operation in  (\ref{eq:splitted-Eq}), $\bar{\bf u}$ should be transformed to ${\bf u}$ through the operation ${\bf u}=\hat{M^\dagger}\bar{\bf u}$ (${\bf u}=\hat{W}\bar{\bf u}$) while using Gaussian (Popov or Lebedev) quadratures (cubatures).

It should be emphasized that both transformation matrices (either $\hat{M}$ or $\hat{W}$) are independent of $r$ and $t$.

For approximating the Schr\"odinger equation (\ref{eq:Sch-type_Eq}) over the radial variable $r$ the six-order finite-difference approximation is used on a quasi-uniform grid $\{r_j\}_1^{N_r}$ on the interval $r\in [0,r_m]$ constructed by mapping the $r$ variable to $x\in [0,1]$ with the formula $r=r_m[\exp(4x)-1]/[\exp(4)-1]$. To avoid the artificial reflection of the electron at the radial grid boundary, the wave function is multiplied after each time step by the absorbing ``mask function'' suggested in Ref.\cite{Krause92}. The total number of operations at each time step (\ref{eq:splitted-Eq}) in our algorithm is $N_r(2N_\Omega^2 + \alpha N_\Omega)$ where $\alpha = (2\times7)^2 \simeq 200$ is given by the width of the band of the band matrices $\hat{A}$ in (\ref{eq:matrixA}) and (\ref{eq:matrixA2}) with the finite-difference approximation of the radial derivatives with respect to the variable $r$. Here we have an obvious advantage with respect the Tong computational scheme \cite{Tong2017} for numerical integration of the Shr\"odinger equation (1) where the number of operations at each time step is $N_r N_\theta N_\phi (N_r+N_\theta+N_\phi)$, i.e. quadratically dependent on the number of radial grid points $N_r$. The computational time of our scheme linearly depends on $N_r$ and  $\sim N_\Omega^\gamma$ with $1\lesssim\gamma<2$  approaching 1 for not very big $N_\Omega$ \cite{Melezhik2002}. Linear dependence of the computational time on $N_r$ permits computations on the radial grids up to $r_m\ = 1000$ and $\Delta x =1/N_r=0.00025$. The time step of integration is chosen to be $\Delta t=0.05$ ($1/2200$ of one optical cycle).

Once the wavefunction $\psi({\bf r},t)$ is calculated, one can investigate the physical properties of the atom in the laser field such as photoelectron momentum spectra, photoexcitation or HHG etc. Here we focus on the ionization and excitation probabilities.
\subsection{Ionization and Excitation}\label{Subsec:Ionization-Theory}
We calculate the transition probabilities in the hydrogen atom due to interaction with a laser pulse by projecting the electron wave-packet at the end of the pulse $\psi({\bf r},t = n_TT)$ on the bound
\begin{eqnarray}
\label{eq:Pnlm}
P_{nlm} = |\langle \phi_{nlm}|\psi\rangle|^2 = |d_{nlm}|^2
\end{eqnarray}
and continuum
\begin{eqnarray}
\label{eq:Pk}
\frac{dP_{lm}}{dE}(E)=|\langle \phi_{klm}|\psi\rangle|^2 =|c_{klm}|^2
\end{eqnarray}
states of the atomic spectrum in free space. The hydrogen eigenfunctions in the above formulas (\ref{eq:Pnlm}) and (\ref{eq:Pk}) of the bound
\begin{eqnarray}
\label{eq:boundF}
\phi_{nlm}({\bf r})=R_{nl}(r)Y_{lm}(\Omega)
\end{eqnarray}
and continuum
\begin{eqnarray}
\label{eq:coulombk}
\phi_{klm}({\bf r})=\sqrt{\frac{2}{\pi k}}\frac{F_l(\eta,kr)}{r}Y_{lm}(\Omega)
\end{eqnarray}
spectrum form an orthonormal and complete basis set which can be used for expansion of the electron wave-packet $\psi({\bf r},t)$ after interaction with the laser pulse
\begin{eqnarray}
\label{eq:expandedPsi}
\psi({\bf r},t = n_TT) = \sum_{nlm}d_{nlm}\phi_{nlm}({\bf r}) + \sum_{lm} \int_0^{+\infty}c_{klm}\phi_{klm}({\bf r})dE \,\,,
\end{eqnarray}
where $E$ denotes the photoelectron energy, $k=\sqrt{2E}$, $\eta=-1/k$ and $F_l(\eta,kr)$ is the Coulomb wavefunction regular as $kr\rightarrow 0$ .

Thus, the population of the $n$-th bound state and the ATI spectrum of the hydrogen atom after interaction with the laser pulse are described as
\begin{eqnarray}
\label{eq:Pn}
P_n=\sum_{l=0}^{n-1}\sum_{m=-l}^{l}P_{nlm}
\end{eqnarray}
and
\begin{eqnarray}
\label{eq:dPdk}
\frac{dP}{dE}(E)=\sum_{l=0}^{l_m\rightarrow\infty}\sum_{m=-l}^{l}\frac{dP_{lm}}{dE}(E)\,.
\end{eqnarray}

Together with the above equations we also use the relation between the
population probabilities
\begin{eqnarray}
\label{eq:sumP}
\sum_{n=1}^{\infty}P_{n} + \int_{0}^{+\infty}\frac{dP}{dE}dE = \langle \psi|\psi\rangle \approx 1\,.
\end{eqnarray}
Due to the action of the absorbing ``mask function'' at the edge of radial integration $r_m$, in our computation we have to take into account the effect that the normalization integral $\langle \psi|\psi\rangle$ decreases in time and deviates from unit. Then the total ionization yield is obtained by
\begin{eqnarray}
\label{eq:Pion}
P_{ion}=\int_{0}^{+\infty}\frac{dP}{dE}dE+\left(1-\langle \psi|\psi\rangle\right)\,,
\end{eqnarray}
and $\sum_{n=1}^{\infty}P_{n}=P_g+P_{ex}$, where $P_g$ and $P_{ex}$ denote the ground state population and total excitation probability.

To circumvent the seemingly intractable problem of calculating the  populations $P_n$ as $n\rightarrow +\infty$ , we propose an alternative method which does not require numerical analysis for very large $n,l,m$ values. It is based on the known property of the Coulomb wave functions $\phi_{nlm}({\bf r})$ (\ref{eq:boundF}) and $\phi_{klm}({\bf r})$ (\ref{eq:coulombk}): the functions $n^{3/2}\phi_{nlm}({\bf r})$ and $\phi_{klm}({\bf r})$ transform one to another as $n\rightarrow +\infty$ and $k\rightarrow 0$ with the replacement $n \leftrightarrow 1/(ik)$ \cite{Bethe1977,Fock1978}.

In this method, we stop calculating $P_n$ at some rather large $n=N^\prime$
\begin{eqnarray}
\label{eq:PexReplaced}
P_{ex}=\sum_{n=2}^{\infty}P_n=\sum_{n=2}^{N^\prime}P_n+\sum_{n=N^\prime+1}^{\infty}P_n
\end{eqnarray}
and evaluate the remaining sum $\sum_{n=N^\prime+1}^{\infty}P_n$ in the above summation according to
\begin{eqnarray}
\label{eq:discardedPn}
\fl\sum_{n=N^\prime+1}^{+\infty}P_n \simeq \int_{n=N^\prime+1}^{+\infty}P_n dn = \int_{N^\prime+1}^{+\infty}n^3P_nd(-\frac{1}{2n^2})=\int_{E^\prime}^0\frac{dP}{dE}(E<0)dE\,,
\end{eqnarray}
where $E^\prime=-\frac{1}{2(N^\prime+1)^2}$ , and $n^3P_n = \frac{dP}{dE}(E<0)$. It is well known \cite{Fano,Vinitskii} (see also \cite{Gao2019}) that in the limit $n\rightarrow +\infty$ the oscillator strength density $\frac{dP}{dE}(E<0)$ approaches the value $\frac{dP}{dE}(-0)$ coinciding with the limit $\frac{dP}{dE}(+0)$ of the population of the states of the continuum spectrum at $ k\rightarrow 0 (E\rightarrow +0)$.

Thus, by calculating $\frac{dP}{dE}(E>0)$ from Eq.(\ref{eq:dPdk}) at some positive small values of $E~(E\rightarrow +0)$, where small $l,m$ give a principal contribution, we can find a trend curve $f(E)=\frac{dP}{dE}(0)+AE+BE^2+...$ passing through these points and the points $n^3P_n=\frac{dP}{dE}(E<0)$ calculated on the negative side of $E$ which belong to the $n=N^\prime,~N^\prime-1,...$ states.
With the calculated $f(E)$ we perform infinite summation in (\ref{eq:discardedPn})
\begin{eqnarray}
\label{eq:FinaldiscardedPn}
\sum_{n=N^\prime+1}^{+\infty}P_n =\sum_{n=N^\prime+1}^{n_{m}}\frac{f(E_n)}{n^3}+\int_{E_1}^0f(E)dE
\end{eqnarray}
where $E_1=E_{n_{m}+1}$ and $n_m$ is chosen to be as large as $30$ after which the difference between summation over $n$ and integration over $E$ becomes negligible. Finally, we can obtain $P_{ex}$ by Eq.(\ref{eq:PexReplaced}) and then the total ionization yield by $P_{ion}=1-P_g-P_{ex}$.

\section{Results and Discussion}
\label{Sec:Results}
Based on the computational schemes presented above, we investigate the dynamics of the hydrogen atom in an elliptically polarized  laser field for different ellipticities $0\leq\epsilon\leq 1$ with the laser intensity of $I=10^{14}$ W/{cm}$^2$ and wavelength of $\lambda=800$ nm for the laser pulse duration of $20$ fs ($=7.5\times T$).

\subsection{Linearly Polarized Laser Field}
\label{Subsec:Linear}

\begin{figure}[ht]
\centering
\includegraphics[width=0.7\textwidth]{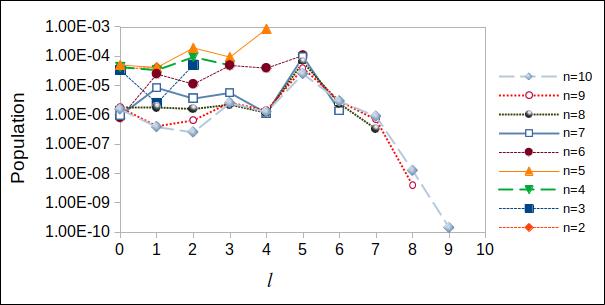}
\caption{The distribution of the population $P_{nl}$ of the excited states of the hydrogen atom while the laser field is linearly polarized along the $z$-axis. The calculations are performed with 1D npDVR Gaussian scheme with $N_\Omega=95$. The vertical axis is in logarithmic scale.}
\label{fig:Fig0}
\end{figure}

\begin{table}[b]
\caption{The transition probabilities $P_n$ to the bound states $n$ of the hydrogen atom calculated with the 1D DVR Gaussian scheme in the case of atomic interaction with the linearly polarized laser field along the $z$-axis. Here, $N_\Omega=N_{\theta}$ , and the contributing $l$ values in the basis functions are  $l=0,1,..,N_\Omega-1$ ($m=0$).}
\label{tab:Table1}
\begin{indented}
\item[]\begin{tabular}{|c|c|c|c|c|c|}
\br
 & \multicolumn{5}{c|}{$P_n$}\\[0.5ex]
\hline 
$N_\Omega$ & 35 & 50 & 65 & 80 & 95\\[0.5ex]
\hline
$n=1$ & 0.984693277 & 0.984677192 &	0.984678262 & 0.984678290 &	0.984678289\\[1ex]
$n=2$ & 8.16E-05 & 7.89E-05 & 7.87E-05 &	7.87E-05 & 7.87E-05\\[1ex]
$n=3$ & 8.83E-05	 & 8.74E-05 & 8.69E-05 & 8.69E-05 & 8.69E-05\\[1ex]
$n=4$ & 2.20E-04 & 2.20E-04 & 2.19E-04 &	2.19E-04 & 2.19E-04\\[1ex]
$n=5$ & 1.23E-03 & 1.24E-03 & 1.24E-03 &	1.24E-03 & 1.24E-03\\[1ex]
$n=6$ & 2.46E-04	 & 2.34E-04 & 2.30E-04 & 2.31E-04 & 2.31E-04\\[1ex]
$n=7$ & 1.27E-04 & 1.08E-04 & 1.18E-04 &	1.17E-04 & 1.17E-04\\[1ex]
$n=8$ & 8.15E-05	 & 7.14E-05 & 7.48E-05 &	7.37E-05 & 7.41E-05\\[1ex]
$n=9$ & 5.45E-05 & 4.95E-05 & 4.95E-05 &	4.94E-05 & 4.94E-05\\[1ex]
$n=10$ & 3.79E-05 & 3.54E-05 & 3.44E-05 & 3.46E-05 &	3.46E-05\\[1ex]
\hline
\end{tabular}
\end{indented}
\end{table}
First, we consider a hydrogen atom in a laser field of linear polarization ($\epsilon=0$). This case permits the separation of the azimuthal angle $\phi$ if the electric field is oriented along the $z$-axis and the 2D DVR basis constructed with the Gaussian scheme (\ref{eq:fj_Gaussian}) reduces to the 1D DVR over the $\theta$ variable with $N_\Omega=N_\theta$ ($l=0,1,..,N_\Omega-1$, and $m=0$). Such simplification of the problem gives us the possibility to get the most accurate result and use it as a test for the convergence of the 2D npDVRs constructed with the Popov and Lebedev cubatures where the separation of angular variables are not allowed, regardless of the chosen orientation for the electric field in space.
In Table \ref{tab:Table1}, we demonstrate the convergence of the scheme based on the Gaussian 1D DVR with respect to the number of the basis function $N_\Omega$  in calculating the populations $P_n$ of the ground ($n=1$) and excited states up to $n=10$. It is shown that including $N_\Omega=35$ basis functions in the 1D DVR Gaussian scheme (where $l \leq 34$ and $m=0$) fixes four significant digits after the decimal point in $P_1$ and gives the relative accuracy $\sim 9\%$ in calculating $P_n$ for $2\leq n \leq 10$. When increasing the number of basis functions to $N_\Omega = 95$ ($l \leq 94$, $m=0$), eight significant digits are fixed in $P_1$, and the relative accuracy reaches the value $\sim 0.5\%$ in $P_n$ for $2\leq n \leq 10$. The distribution of the populations $P_{nl}$ over the angular momentum $l$ is presented in Fig.\ref{fig:Fig0}. Here, the most probable angular momentum states are observed with $l\leqslant 5$.

\begin{figure}[t]
\centering
\includegraphics[width=0.7\textwidth]{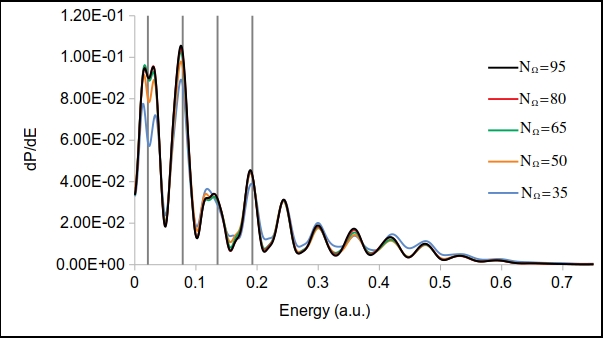}
\caption{ (Color online) Energy distribution of electrons $\frac{dP}{dE}(E)$ due to ionization of the hydrogen atom by a laser field linearly polarized along the $z$-axis. The $\frac{dP}{dE}(E)$ distribution is calculated according to the 1D DVR Gaussian scheme as a function of $N_\Omega$. Vertical gray lines correspond to the expected peaks of the multiphoton ionization process in accordance with Eq.(\ref{eq:Epeak-position}). }
\label{fig:Fig1}
\end{figure}

In Fig.\ref{fig:Fig1}, we demonstrate the convergence $N_\Omega \rightarrow +\infty$ of the transition probabilities $\frac{dP}{dE}(E)$ to the states of the continuum spectra of the hydrogen atom due to interaction with a laser pulse linearly polarized along the $z$-axis. The calculations are performed with the 1D DVR Gaussian scheme where the $\phi$-variable is separated and $m=0$. In the calculations, the highest angular momentum in Eq.(\ref{eq:dPdk}) was fixed at $l_{m} = 35$ by the requirement that the error in cutting off summation (\ref{eq:dPdk}) does not exceed the error in using the limited DVR basis. The spectrum contains several peaks whose position does not depend on the number of basis functions $N_\Omega$ and are almost placed at the peak positions $E_p$ of the energy spectrum in a multiphoton process
\cite{Telnov,Piraux2010,ESpectra2011}
\begin{eqnarray}\label{eq:Epeak-position}
E_p=p\hbar\omega-U_{pe}-I_i\,\,,
\end{eqnarray}
where $I_i=13.6$ eV is the ionization energy of the hydrogen atom, $\hbar\omega=1.55$ eV is the photon energy, and $U_{pe}=I/\left(4\omega^2\right)=5.967$ eV is the ponderomotive energy of our laser field with intensity $I=10^{14}$ W/{cm}$^2$. The calculated $\frac{dP}{dE}(E)$ is qualitatively in agreement with Ref.\cite{Piraux2010} where a $25$ fs laser pulse was used. For this high intensity, the first expected peak corresponding to the 13-photon ionization, $p=13$ in Eq.(\ref{eq:Epeak-position}), is splitted, just as in \cite{Piraux2010}. The various mechanisms leading to the doubling of the first ATI peak was intensively discussed earlier \cite{Photoabsorp1,Milosevic2006, Dimitr2004, Nganso2011}. In the considered case, apparently, the intermediate resonant population of some of the levels dressed by strong laser field is responsible for this effect \cite{Photoabsorp1}. However, to clarify this issue, a detailed calculation of the level deformations by a strong laser pulse is required.

\begin{figure}[b]
\centering
\includegraphics[width=0.7\textwidth]{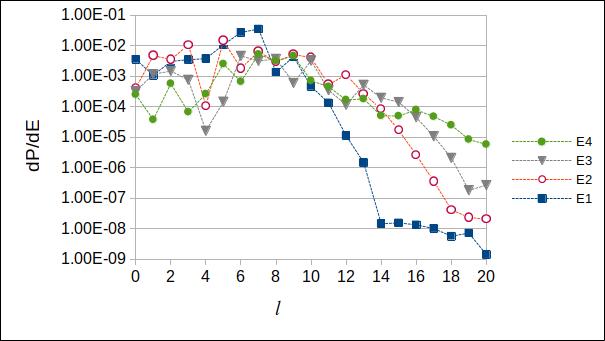}
\caption{ (Color online) The distribution of electrons $\frac{dP_l}{dE}(E)=\sum_{m=-l}^{l}\frac{dP_{lm}}{dE}(E)$ over angluar momentum $l$ due to ionization of the hydrogen atom by a laser field linearly polarized along the $z$-axis (thus $m=0$). Four different electron energy values $E$ above successive peaks $E=E_p+0.01$ (in a.u.) are considered where $E_p$ is obtained according to (\ref{eq:Epeak-position}) and $p=13,14,15,16$ corresponds to E1 (blue squares), E2 (open red circles), E3 (gray triangles) and E4 (full green circles). Here, the calculations are performed with $N_\Omega=95$. The vertical axis is in logarithmic scale.}
\label{fig:Fig11}
\end{figure}

The distribution $\frac{dP_l}{dE}(E)$ over electron angular momentum $l$ in the continuum for different energies $E$ between successive peaks ($E=E_p+0.01$ (in a.u.), where $p$ corresponds to $13,14,15,16-$photon ionization) is presented in Fig.\ref{fig:Fig11}. As can be seen in this figure, by increasing the electron energy, the contribution of higher $l$-values to the continuous spectrum is increased. However, the main contribution comes from angular momenta $0\leqslant l \leqslant 12$.
\begin{table}[t]
\caption{The ionization yield $P_{ion}$ and the total excitation probability $P_{ex}$ as a function of  $N_\Omega$ in the Gaussian 1D DVR approach for the linearly polarized laser field along the $z$-axis. The $P_{ion}$ values are calculated by integrating the transition probabilities $\frac{dP}{dE}(E)$ of the electron continuum over $E$ through Eqs.(\ref{eq:Pk},\ref{eq:dPdk},\ref{eq:Pion}) and the excitation probabilities $P_{ex}$ by $P_{ex}=1-P_g-P_{ion}$.} 
\label{tab:Table2}
\begin{indented}
\item[]\begin{tabular}{|c|ccccc|}
\br
$N_\Omega$ & 35&	50&	65&	80&	95\\ [0.5ex]
\hline 
$P_{ion}$ & 1.298E-02 &	1.284E-02 &	1.301E-02 &	1.304E-02 &	1.304E-02\\[1ex] 
$P_{ex}$  & 2.330E-03 &	2.484E-03 &	2.311E-03 &	2.286E-03 &	2.284E-03\\[1ex]
\hline 
\end{tabular}
\end{indented}
\end{table}
\begin{figure}[b]
\centering\includegraphics[width=0.6\textwidth]{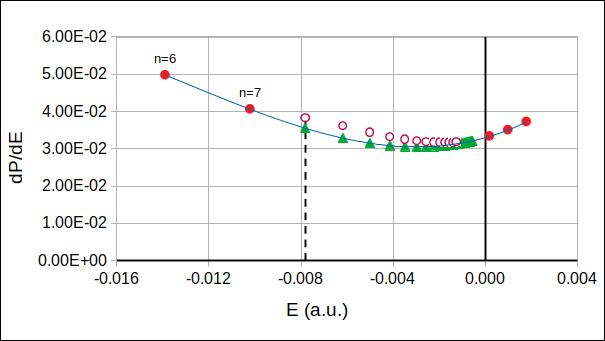}
\caption{(Color online) Illustration of the $P_{ex}$ calculation by formulas (\ref{eq:PexReplaced}) and (\ref{eq:FinaldiscardedPn}) in the case of linearly polarized laser field along the $z$-axis. Two full circles $\frac{dP}{dE}(E_n)$ on the negative part of the electron energy $E_n=-\frac{1}{2n^2}$ are calculated for $n=6$, $7$ by the relation $\frac{dP}{dE}(E_n)=n^3P_n$. Whereas for three full circles on the positive side of $E$, we use the projection of the calculated electron wave-packet at the end of the laser pulse onto Coulomb states of the continuum spectrum via Eqs.(\ref{eq:Pk},\ref{eq:dPdk}). A cubic polynomial trend line (blue line) $f(E)$ passes through these circles. The open circles $\frac{dP}{dE}(E_n)$ belong to the values calculated by the Gaussian 1D DVR scheme for $n=8,9,...,20$ and the green triangles show the corresponding interpolated $f(E_n)$. The calculations are performed within the Gaussian 1D DVR for $N_\Omega=65$.}
\label{fig:Fig2}
\end{figure}

In Table \ref{tab:Table2}, the total ionization yield $P_{ion}$ which is calculated by integrating the transition probabilities $\frac{dP}{dE}(E)$ over all possible states of the ionized electron in its continuum (\ref{eq:Pion}), and the total excitation probability $P_{ex}=1-P_g-P_{ion}$ are presented. These results demonstrate the convergence of the computational scheme in calculating $ P_ {ion} $ and $ P_ {ex} $ at the accuracy level of the order of the fourth significant digit after the decimal point.

The total excitation probability $P_{ex}$ can also be calculated by direct summation of $P_{nl}$ over $n$ and $l$. However, this requires taking into account the contribution of large $n$ and $l$ values which in some cases can be considerable \cite{Photoabsorp2, Fedorov1988, Fedorov1997} and greatly complicates the problem computationally. Therefore, we propose an alternative method described in the previous section , equations (\ref{eq:PexReplaced}) and (\ref{eq:FinaldiscardedPn}), to calculate $P_{ex}$ without summation over large $n$ and $l$. To get the probability $P_{ex}$ according to these formulas (the procedure is illustrated in Fig.\ref{fig:Fig2}), we have calculated the populations $P_n$ of the hydrogen atom bound states up to $n=7$ , according to (\ref{eq:Pn}), and followed the method described in Sec.\ref{Subsec:Ionization-Theory} for calculating the populations of states with $n\geq 8$. The points $\frac{dP}{dE}(E_n)=n^3P_n$ on the negative side of the electron energy in Fig.\ref{fig:Fig2} belong to $n=6$ and $n=7$ states. On the positive side, three $E$ points were chosen very close to the zero energy where $\frac{dP}{dE}(E)$ were calculated by Eqs.(\ref{eq:Pk},\ref{eq:dPdk}). By these five points the polynomial function $f(E)$ was fixed for calculating $\sum_{n=8}^{+\infty}P_n$ , according to the relation (\ref{eq:FinaldiscardedPn}).
In Fig.\ref{fig:Fig2} we see a good agreement between the calculated curve $f(E)$ and the corresponding values calculated for $n$ up to $n=20$ by the direct way (\ref{eq:Pnlm}). It confirms that the choice of $n=6$ and $n=7$ when constructing $f(E)$ gives a good approximation which can be used to evaluate $P_n$ for large $n$.
The probabilities $P_{ex}$ calculated with this method and $P_{ion}=1-P_g-P_{ex}$ as a function of $N_\Omega$ are presented in Table \ref{tab:Table3}. By comparing the results presented in Tables \ref{tab:Table2} and \ref{tab:Table3}, one can conclude that our method (\ref{eq:PexReplaced},\ref{eq:FinaldiscardedPn}) proposed for calculating the total excitation probability $P_{ex}$ and ionization yield $P_{ion}$ generates values which coincide, within a given accuracy, with the corresponding values obtained above by integrating the electron transition probabilities over the states of an electron in the continuum spectrum.

\begin{table}[t]
\caption{The total excitation probability $P_{ex}$ and the ionization yield $P_{ion}$ as a function of $N_\Omega$ within the Gaussian 1D DVR scheme in the case of linearly polarized laser field along the $z$-axis. The $P_{ex}$ values are calculated, according to the formulas (\ref{eq:PexReplaced},\ref{eq:FinaldiscardedPn}), approximating the summation over all states of the discrete spectrum of the hydrogen atom (see text).}
\label{tab:Table3}
\begin{indented}
\item[]\begin{tabular}{|c|c|c|c|c|}
\br
$N_\Omega$ & $\sum_{n=2}^{7}P_n$ &	$\sum_{n=8}^{+\infty}P_n$ &	$P_{ex}=\sum_{n=2}^{+\infty}P_n$ &	$P_{ion}=1-P_g-P_{ex}$\\ [0.5ex]
\hline 
35 & 1.994E-03 & 3.047E-04 & 2.299E-03 & 1.301E-02 \\[1ex]
50 & 1.967E-03 & 2.856E-04 & 2.253E-03 & 1.307E-02 \\[1ex]
65 & 1.971E-03 & 2.864E-04 & 2.257E-03 & 1.306E-02 \\[1ex]
80 & 1.971E-03 & 2.882E-04 & 2.259E-03 & 1.306E-02 \\[1ex]
95 & 1.971E-03 & 2.872E-04 & 2.258E-03 & 1.306E-02 \\[1ex]
\hline 
\end{tabular}
\end{indented}
\end{table}

\begin{figure}[t]
\centering\includegraphics[width=0.6\textwidth]{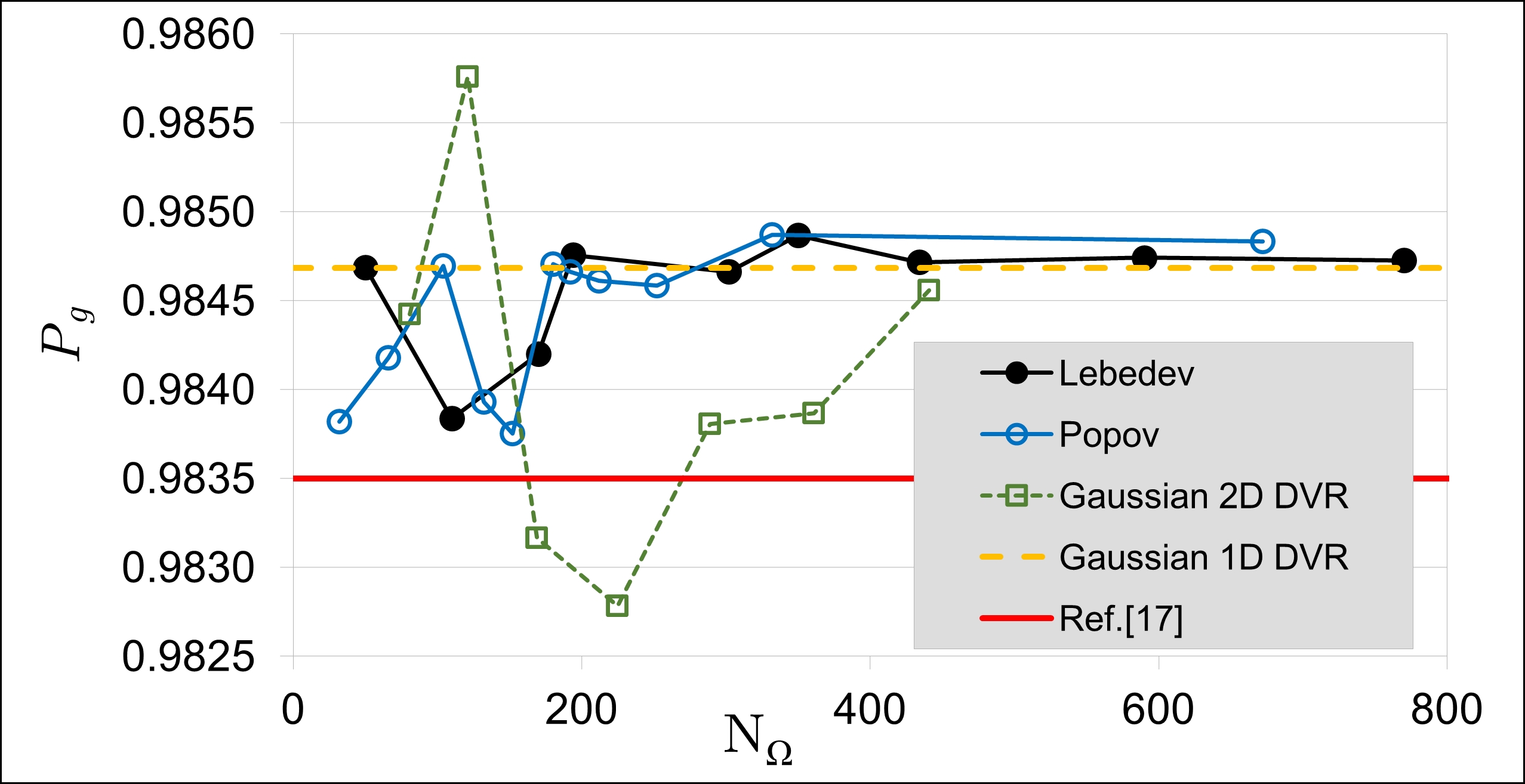}
\caption{(Color online) The population of the ground state  $P_g$, calculated using the 2D npDVR schemes of Lebedev and Popov, as well as the Gaussian 2D and 1D DVR schemes for a hydrogen atom in a linearly polarized laser field, together with the result of Ref.\cite{Gao2019}. The yellow dashed line indicates the most accurate value $P_g$ obtained with the Gaussian 1D DVR scheme for the laser field polarization along the $z-$axis. The open squares indicate the values $P_g$ calculated with the 2D Gaussian DVR for the laser field polarization along the $x-$axis, when the $\phi$ variable is nonseparable and $N_\Omega =N_\theta \times N_\phi$.}
\label{fig:Fig3}
\end{figure}

The values of $P_g$, $P_{ex}$ and $P_{ion}$, calculated above with high accuracy as $N_\Omega=N_\theta \rightarrow +\infty$ in the framework of the Gaussian 1D DVR, hereinafter are used as ``exact'' reference results to investigate the convergence and evaluate the accuracy of the computational schemes based on the Popov and Lebedev 2D npDVRs, where the separation of the $\phi$-variable is not allowed regardless of the laser field orientation. In Fig.\ref{fig:Fig3} , we demonstrate the convergence of the computational schemes based on 2D npDVRs by the example of calculating the population of the ground state $P_g$ as a function of the number of the angular grid points on the unit sphere $N_\Omega$ (the number of basis functions (\ref{eq:fj_Popov}) in 2D npDVR expansion (\ref{eq:main-psi-expansion})). As can be seen in this figure, the ground state population, calculated with the Popov or Lebedev schemes, tends to the ``exact'' value obtained according to the Gaussian scheme (yellow dashed line). We have also performed the calculation of $P_g(N_\Omega)$ for the nonseparable case of the Gaussian 2D DVR,  when the laser field is oriented along the $x$-axis and $N_\Omega=N_\theta\times N_\phi$. Figure \ref{fig:Fig3} displays much faster convergence of the Popov and Lebedev schemes compared to the Gaussian 2D DVR. The slowdown of the latter can be explained by the known effect of clustering of grid points $\Omega_j$ in the Gaussian scheme near the poles $z=0$ as $N_\Omega\rightarrow \infty$ \cite{Beentjes2015} and a more adequate approximation of the angular part of the 3D TDSE in 2D npDVRs.
\begin{table}[b]
\caption{Ground state population $P_g$, total excitation probability $P_{ex}$ and total ionization yield $P_{ion}$ calculated with different 2DnpDVRs and Gaussian 1D DVR together with the results of Ref.\cite{Gao2019} for a hydrogen atom in a linearly polarized laser field. $P_{ex}$ were calculated using the formulas (\ref{eq:PexReplaced},\ref{eq:FinaldiscardedPn}) of the method proposed in Sec.\ref{Subsec:Ionization-Theory} for infinite summation of the populations of the hydrogen bound states and $P_{ion}=1-P_g-P_{ex}$.} 
\label{tab:Table4}
\begin{indented}
\item[]\begin{tabular}{|c|c|c|c|c|}
\br
Methods &  Popov & Lebedev &  Gaussian \footnotesize{(${\bf E}=E{\bf\hat{z}}$)} & Ref.\cite{Gao2019}  \\ [0.5ex]
\hline 
$P_{g}$   &	 0.9848   & 0.9847 & 0.9847   & 0.9835 \\[1ex]
$P_{ex}$  &	 1.90E-03 &  2.15E-03 & 2.26E-03 & 4.18E-03 \\[1ex]
$P_{ion}$ &	 1.33E-02 &  1.32E-02 & 1.31E-02 & 1.23E-02\\[1ex]
\hline 
\end{tabular}
\end{indented}
\end{table}
\begin{table}[t]
\caption{The transition probabilities $P_n$ to the bound states $n$ of the hydrogen atom in terms of increasing $r_m$ and $N_r$ while the ratio $\frac{r_m}{N_r}=0.25$ is fixed. The calculations were performed with Gaussian 1D DVR for $N_\Omega=N_{\theta}=95$ and the linearly polarized laser field along the $z$-axis.}
\label{tab:Table5}
\begin{indented}
\item[]\begin{tabular}{|c|c|c|c|}
\br
 & \multicolumn{3}{c|}{$P_n$}\\ [0.5ex]
\hline 
 & $r_m=250$ , $N_r=1000$ & $r_m=500$ , $N_r=2000$ & $r_m=1000$ , $N_r=4000$ \\[0.5ex]
\hline
$n=1$  & 0.9846820 & 0.9846783 & 0.9846764 \\[1ex]
$n=2$  & 7.87E-05 &	7.87E-05 &	7.87E-05\\[1ex]
$n=3$  & 8.69E-05 &	8.69E-05 &	8.69E-05\\[1ex]
$n=4$  & 2.19E-04 &	2.19E-04 &	2.19E-04\\[1ex]
$n=5$  & 1.24E-03 &	1.24E-03 &	1.24E-03\\[1ex]
$n=6$  & 2.31E-04 &	2.31E-04 &	2.31E-04\\[1ex]
$n=7$  & 1.17E-04 &	1.17E-04 &	1.17E-04\\[1ex]
$n=8$  & 7.40E-05 &	7.41E-05 &	7.41E-05\\[1ex]
$n=9$  & 4.95E-05 &	4.94E-05 &	4.95E-05\\[1ex]
$n=10$ & 3.37E-05 &	3.46E-05 &	3.46E-05\\[1ex]
\hline
$P_{ex}$ & 2.28E-03 & 2.26E-03 & 2.26E-03 \\[1ex]
\hline
$P_{ion}$ & 1.304E-02 & 1.306E-02 & 1.306-02 \\[1ex]
\hline 
\end{tabular}
\end{indented}
\end{table}

A comparison of the different applied schemes in calculating $P_g$, $P_{ex}$ and $P_{ion}$ is summarized in Table \ref{tab:Table4}. Here the largest number of employed Popov and Lebedev grid points is $N_\Omega=672$ and $1202$, respectively. A comparison of the most accurate values obtained with the Lebedev 2D npDVR scheme at $N_\Omega=1202$  with the ``exact'' values calculated using the 1D Gaussian DVR for field polarization along the $z$-axis demonstrates the achieved absolute accuracy of this scheme of the order of $\sim 10^{-5}$ for $P_g$ and $\sim 10^{-4}$ for $P_{ex}$ and $P_{ion}$.
This estimated accuracy also coincides with the value which is obtained through the generally accepted procedure for calculating the error of the numerical methods on a sequence of condensing grids where the error is equal to the attained minimum deviation of the calculated value on two nearest most detailed grids.
Indeed, in our case, for the deviation of the populations calculated on the most detailed grid from its closest number of nodes we get $\Delta P_g = \mid P_g (N_\Omega = 974) -P_g(N_\Omega = 1202)\mid \simeq 2\times 10^{-5}$, $ \Delta P_{ex}, \Delta P_{ion} \simeq 10^{-4}$. This method is also used in the next section for evaluating the accuracy of calculations.
The above results were obtained on a radial grid with $r_m = 500$ and $\Delta x = 0.0005$. To evaluate the error of the approximation over radial variable we have performed calculations with increasing radial box size and decreasing step of integration up to $r_m=1000$ and $\Delta x =1/N_r=0.00025$. The results given in Table \ref{tab:Table5} demonstrate that the approximation error over the radial variable on the grid with $r_m = 500$ and $\Delta x = 0.0005$ is less than due to truncation of the 2D npDVR. Therefore, hereafter we choose this radial grid. The step of integration $\Delta t=0.05$ ($1/2200$ of one optical cycle) over time, which we use, also gives an error that is significantly less than the error introduced by truncating the 2D npDVR.

Although our results deviate from the result of Ref.\cite{Gao2019} (the maximal deviation is for $P_{ex}$ values which differ from one another twice), our three DVR computational schemes converge to the same value in the frame of the reached accuracy. Moreover, the deviation of our results from the values obtained in \cite{Gao2019} exceeds our accuracy by an order of magnitude.

In the next section we apply the 2D npDVRs for a hydrogen atom in elliptically polarized laser fields in the entire range of possible change in ellipticity.

\subsection{Elliptically Polarized Laser Field}
\label{Subsec:Elliptical}
\begin{figure}[ht]
\begin{center}
\begin{subfigure}{
\includegraphics[width=0.4\columnwidth]{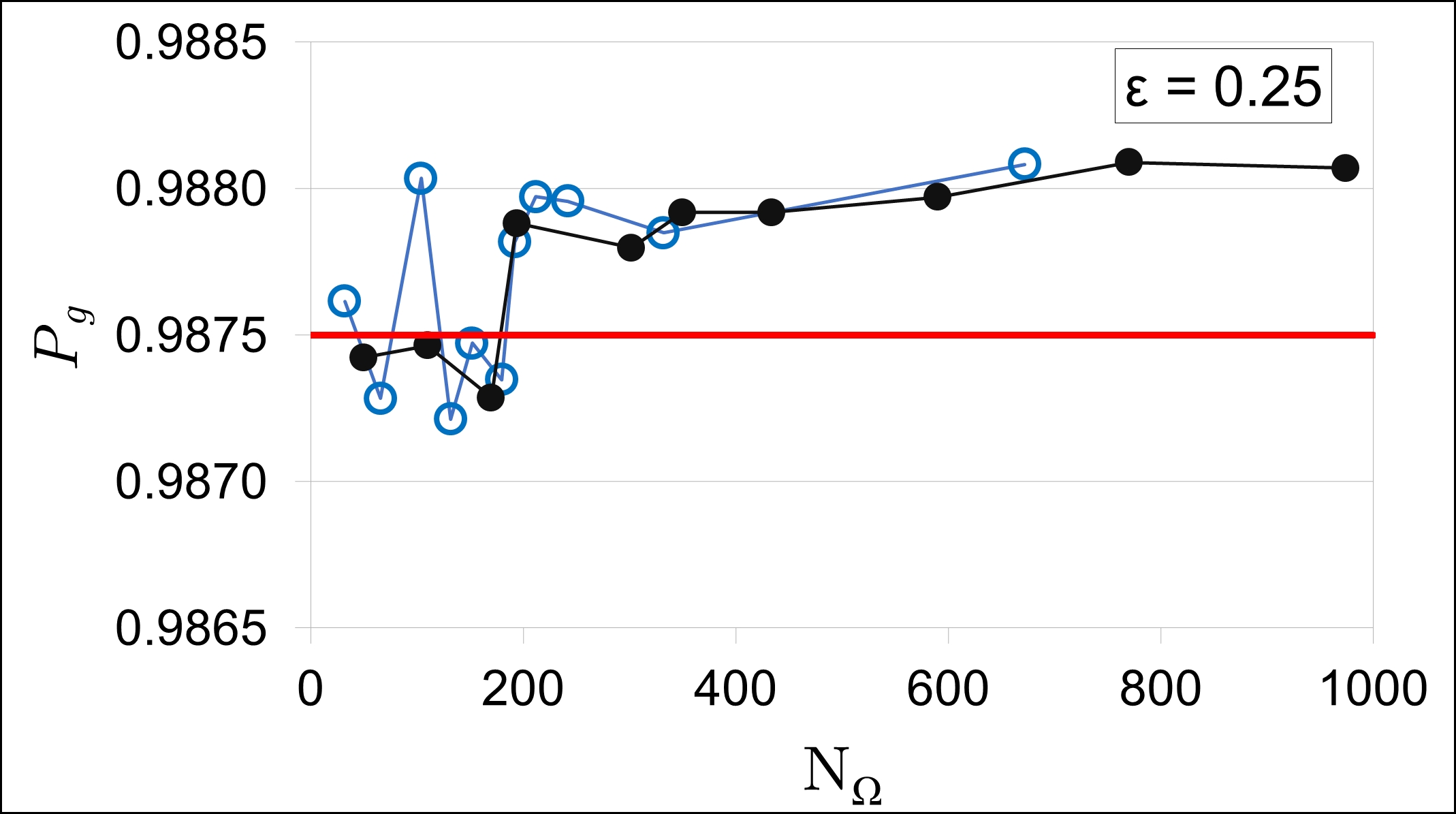} \includegraphics[width=0.4\columnwidth]{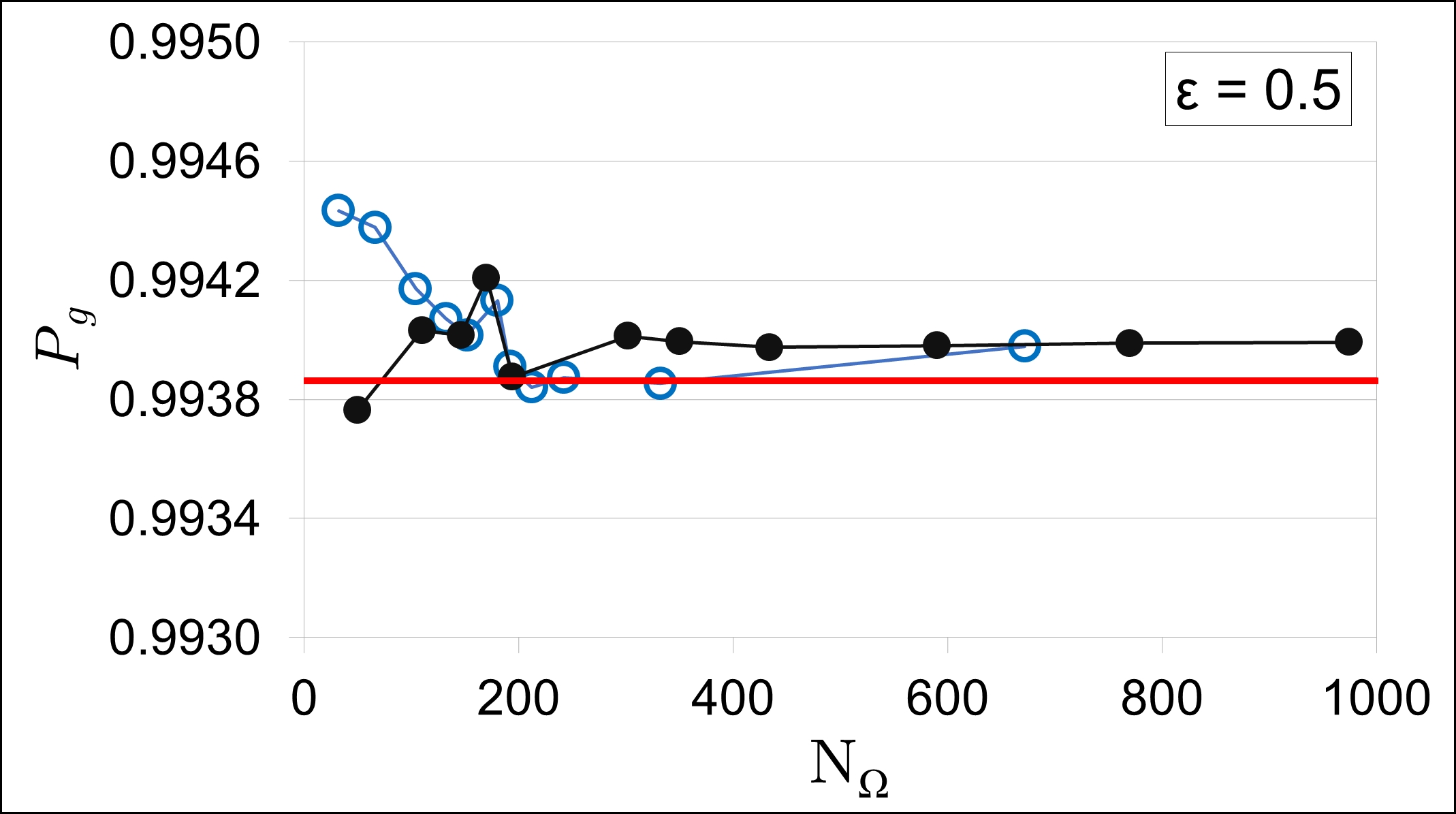}}
\end{subfigure}\\
\begin{subfigure}{
\includegraphics[width=0.4\columnwidth]{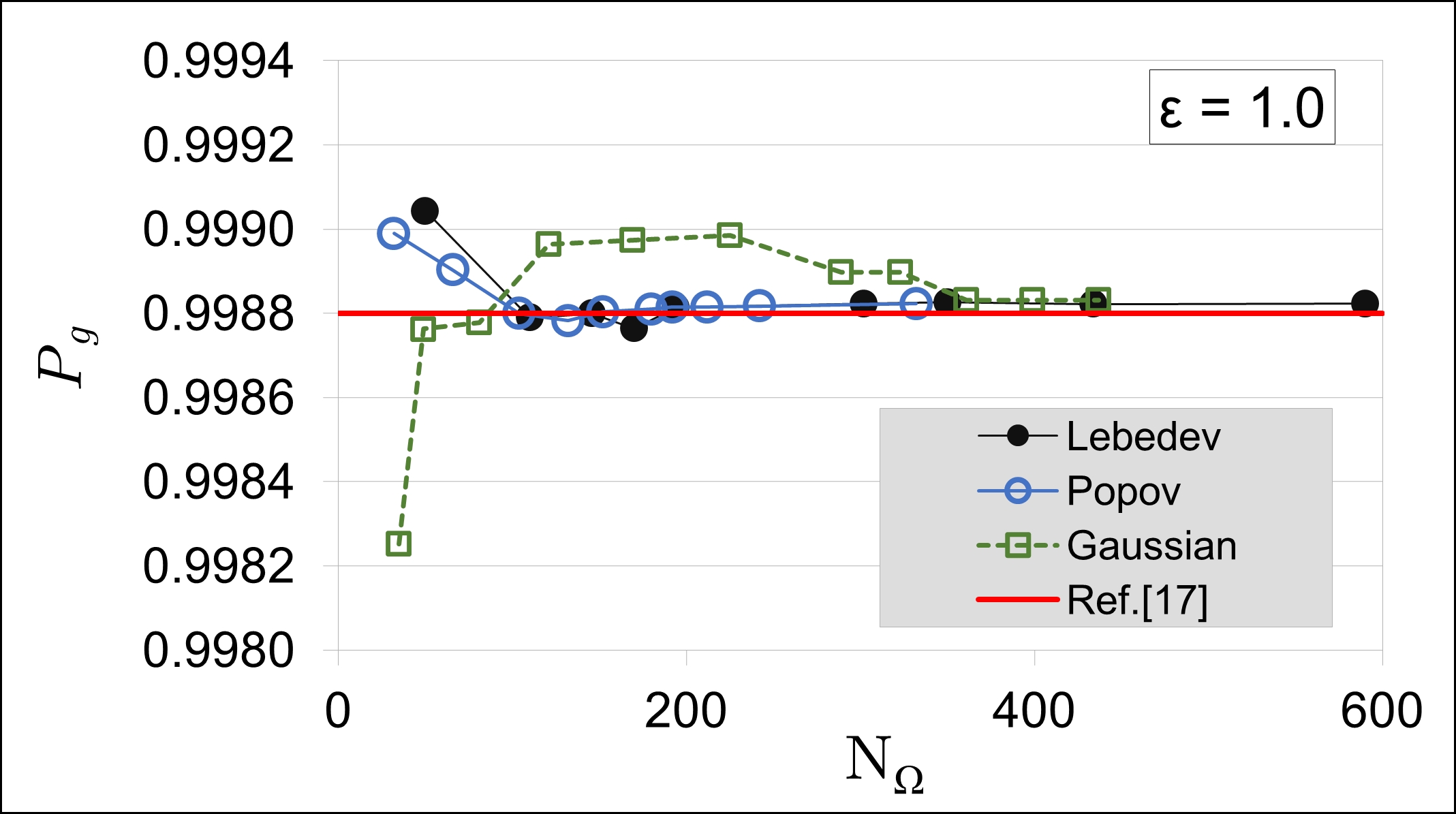}}
\end{subfigure}
\end{center}
\caption{ (Color online) The calculated ground state probabilities by the Popov and Lebedev 2D npDVR schemes for a hydrogen atom in a laser field of different polarizations $\epsilon=0.25$, $0.5$, $1.0$. The red horizontal lines show the corresponding values of Gao and Tong \cite{Gao2019}. For circular polarization, the results obtained with the Gaussian 2D DVR scheme are also presented.}
\label{fig:Fig4}
\end{figure}
\begin{figure}[ht]
\begin{center}
\begin{subfigure}{
\includegraphics[width=0.55\textwidth]{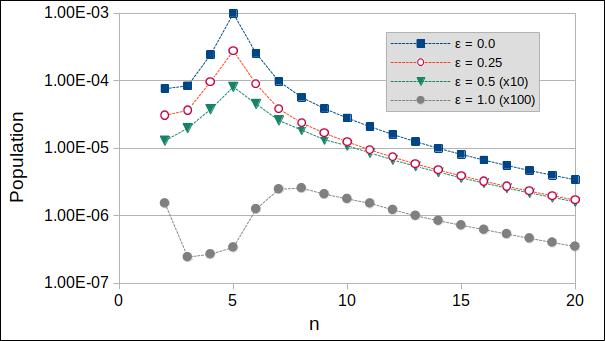}}
\end{subfigure}
\begin{subfigure}{
\includegraphics[width=0.55\textwidth]{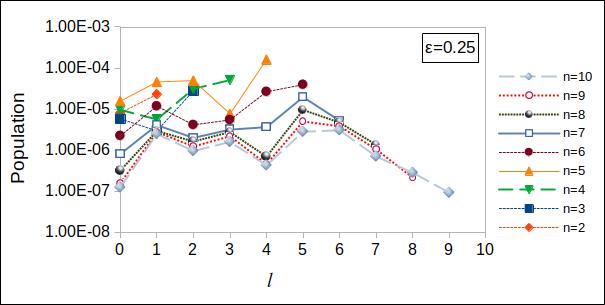}}
\end{subfigure}
\begin{subfigure}{
\includegraphics[width=0.55\textwidth]{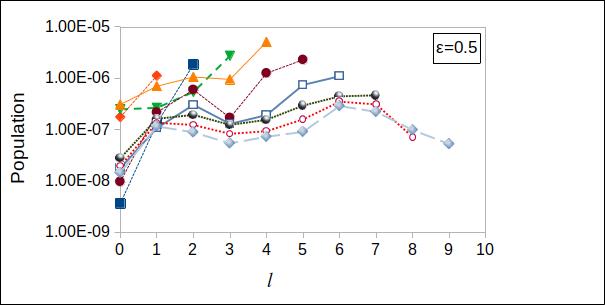}}
\end{subfigure}
\begin{subfigure}{
\includegraphics[width=0.55\textwidth]{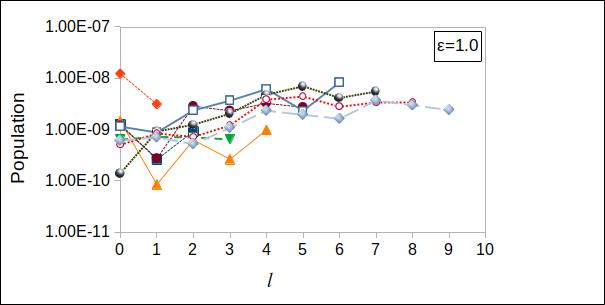}}
\end{subfigure}
\end{center}
\caption{The $n$ and $l$ distribution of the population of excited states for different ellipticities $\epsilon=0.25,0.5,1.0$. The calculations are performed with 2D npDVR Lebedev scheme with $N_\Omega=770$. All vertical axes are in logarithmic scale.}
\label{fig:Fig45}
\end{figure}
The dynamics of a hydrogen atom in a laser field of three different polarizations $\epsilon=0.25$, $0.5$, $1.0$ has been investigated with the developed Popov and Lebedev 2D npDVRs. As was mentioned above, the 1D Gaussian DVR is not applicable for a hydrogen atom in a laser field with $\epsilon\neq 0$. The calculated ground state probabilities $P_g$ are presented in Fig.\ref{fig:Fig4}, which demonstrates the convergence of our 2D DVR computational schemes as $N_\Omega\rightarrow +\infty$. As in the case of linear polarization, the Popov and Lebedev 2D npDVRs schemes demonstrate more rapid convergence compared to the 2D Gaussian DVR for circular polarization as well. As in the case of linear polarization, all three computational schemes converge to the same values, which, however, significantly deviate from the results obtained in \cite{Gao2019}. This deviation decreases with increasing ellipticity and becomes minimal at $\epsilon = 1$.

The distributions of the population ($P_n$ and $P_{nl}$) of excited states over principal quantum number $n$ and angular momentum $l$ are presented in Fig.\ref{fig:Fig45}. While for ellipticities $\epsilon=0.0,0.25,0.5$ a maximum is observed at $n=5$ bound state and a strong decrease in $P_n$ for higher excited states (in agreement with \cite{Photoabsorp2}), the $n$ distribution shows a different trend for the circularly polarized laser field.

\begin{figure}[!]
\begin{center}
\begin{subfigure}{
\includegraphics[width=0.4\columnwidth]{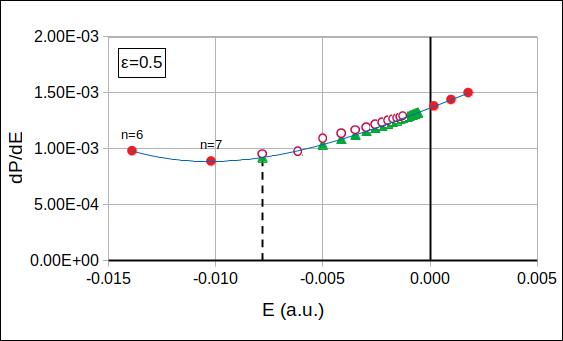} \includegraphics[width=0.4\columnwidth]{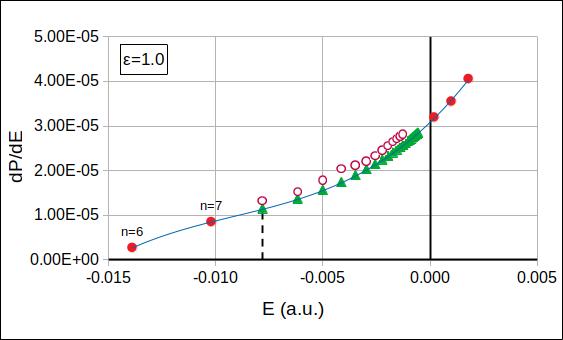}}
\end{subfigure}
\end{center}
\caption{(Color online) Illustration of the $P_{ex}$ calculation by formulas (\ref{eq:PexReplaced}) and (\ref{eq:FinaldiscardedPn}) for the elliptically polarized laser fields with $\epsilon=0.5$, and $1.0$. The calculations of $\frac{dP}{dE}$ are performed with Lebedev 2D npDVR scheme at $N_\Omega=770$. All designations in the figure are similar to those in Fig.\ref{fig:Fig2}.}
\label{fig:Fig5}
\end{figure}

\begin{figure}[ht]
\begin{center}
\begin{subfigure}{
\includegraphics[width=0.4\columnwidth]{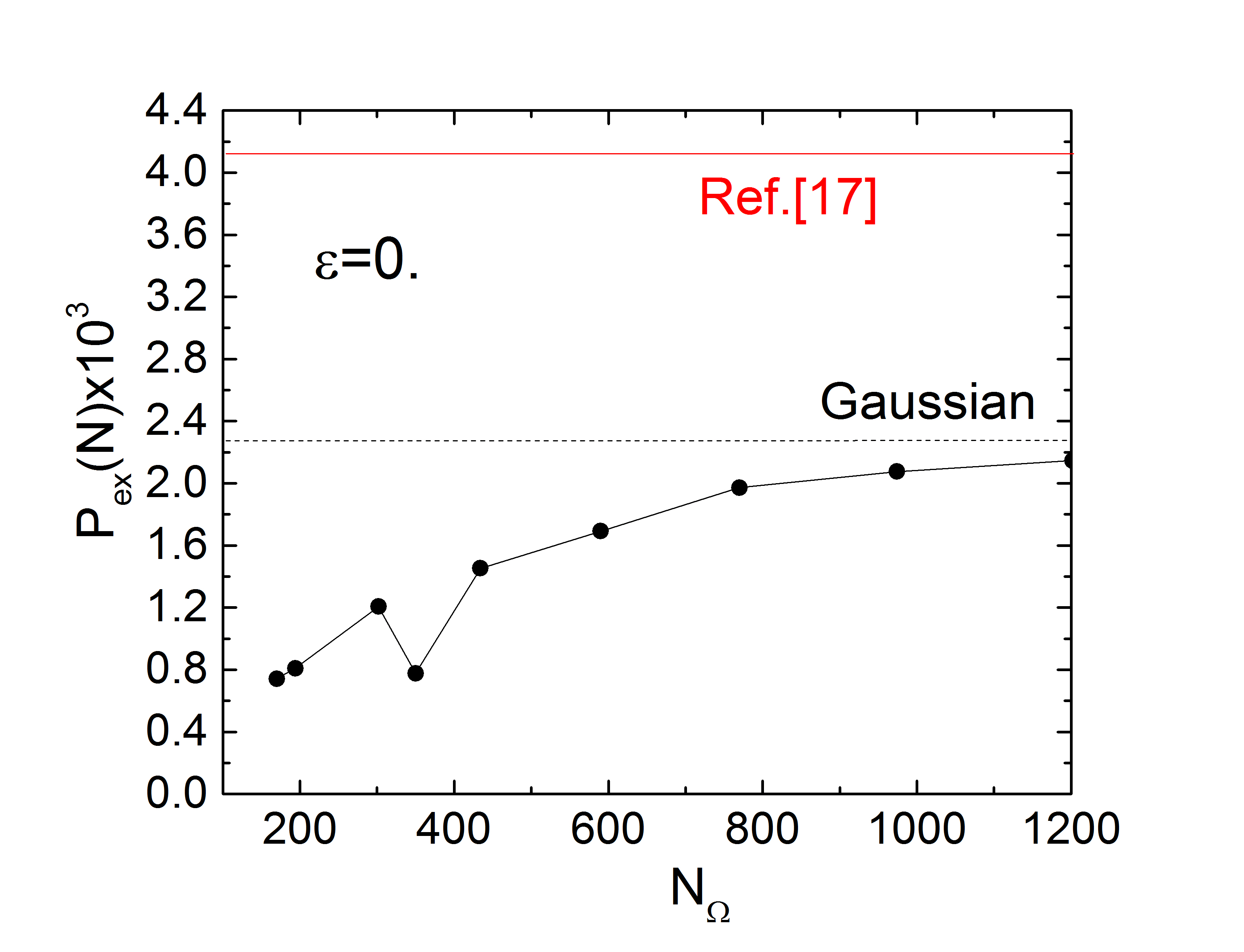} \includegraphics[width=0.4\columnwidth]{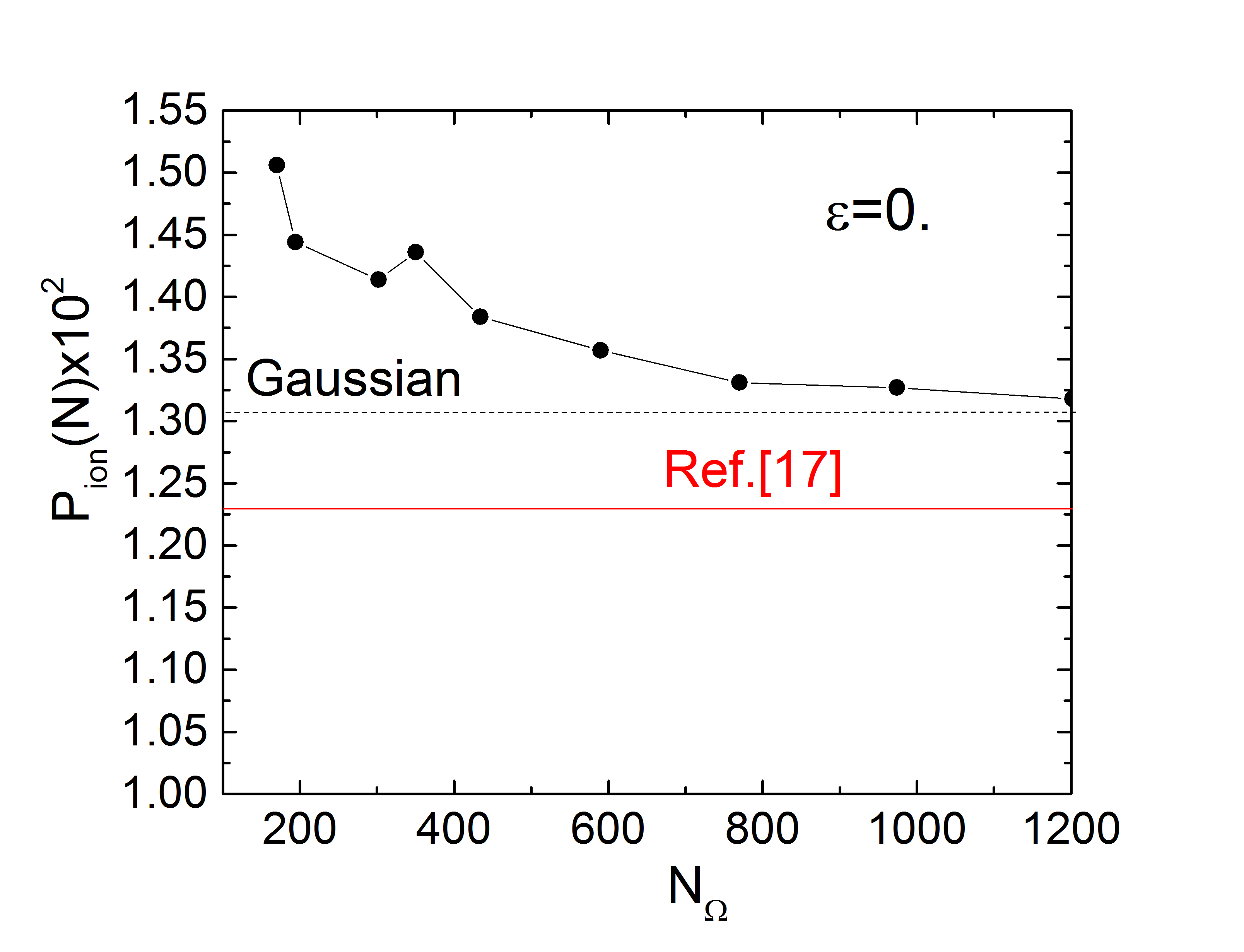}}
\end{subfigure}
\begin{subfigure}{
\includegraphics[width=0.4\columnwidth]{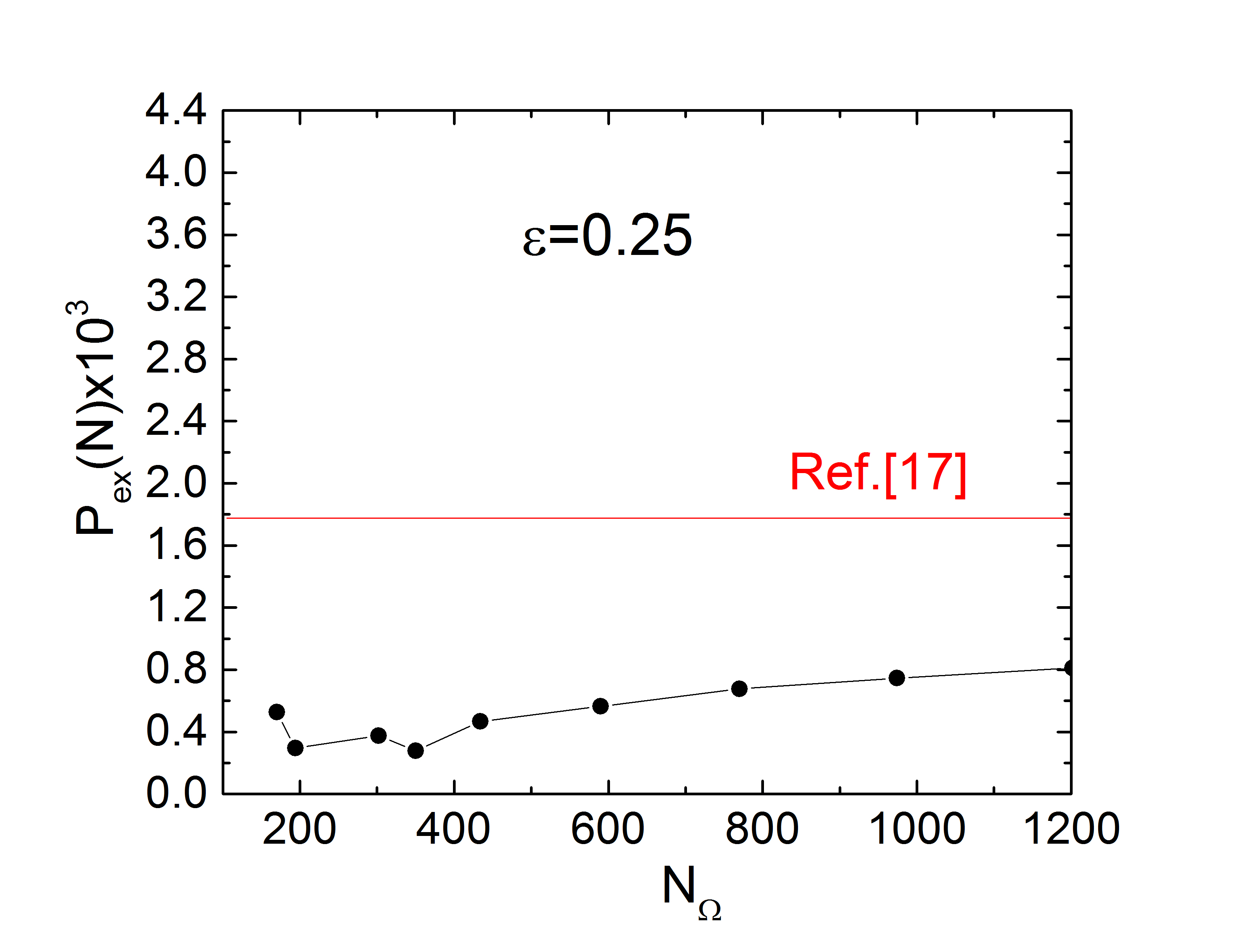} \includegraphics[width=0.4\columnwidth]{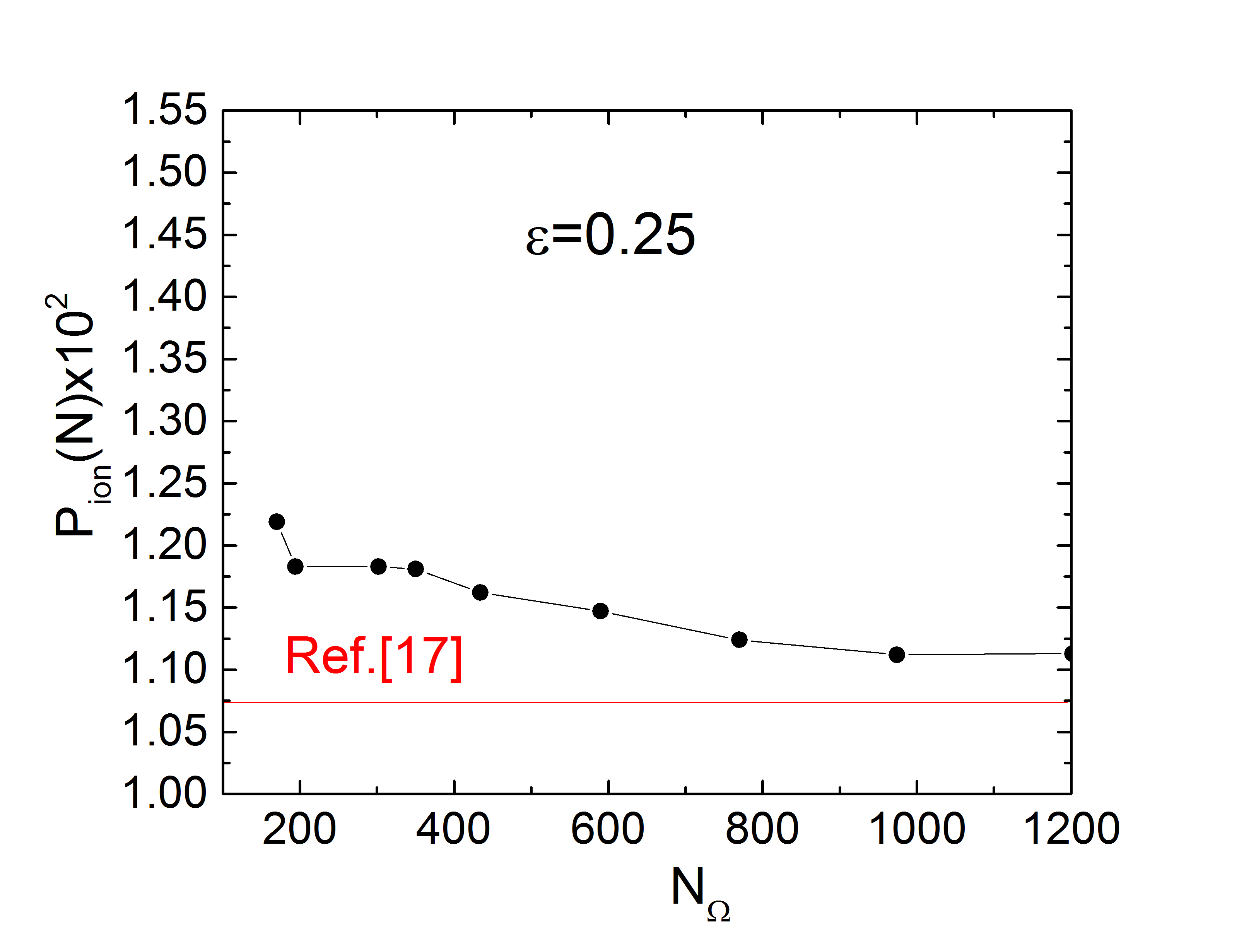}}
\end{subfigure}
\begin{subfigure}{
\includegraphics[width=0.4\columnwidth]{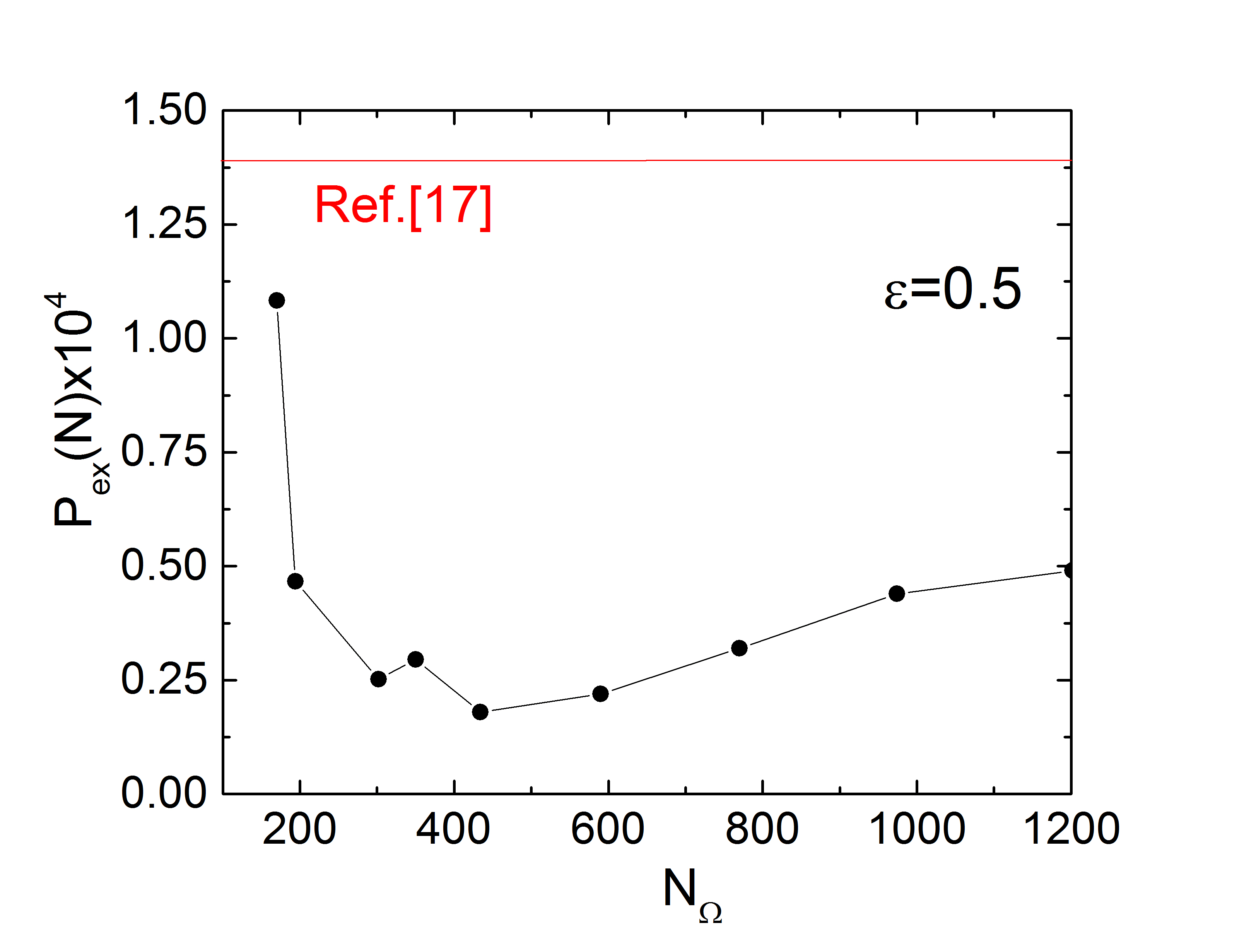} \includegraphics[width=0.4\columnwidth]{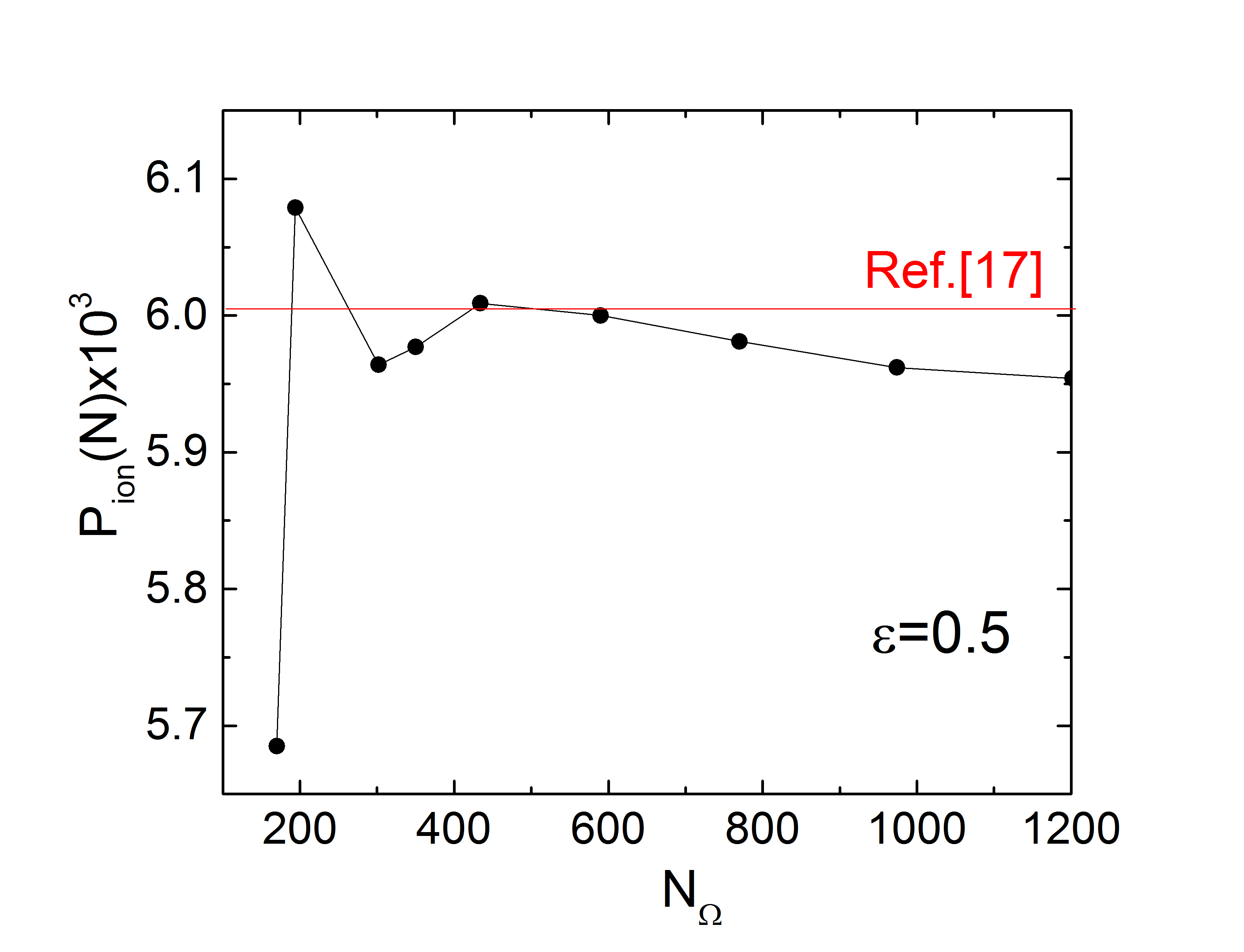}}
\end{subfigure}
\begin{subfigure}{
\includegraphics[width=0.4\columnwidth]{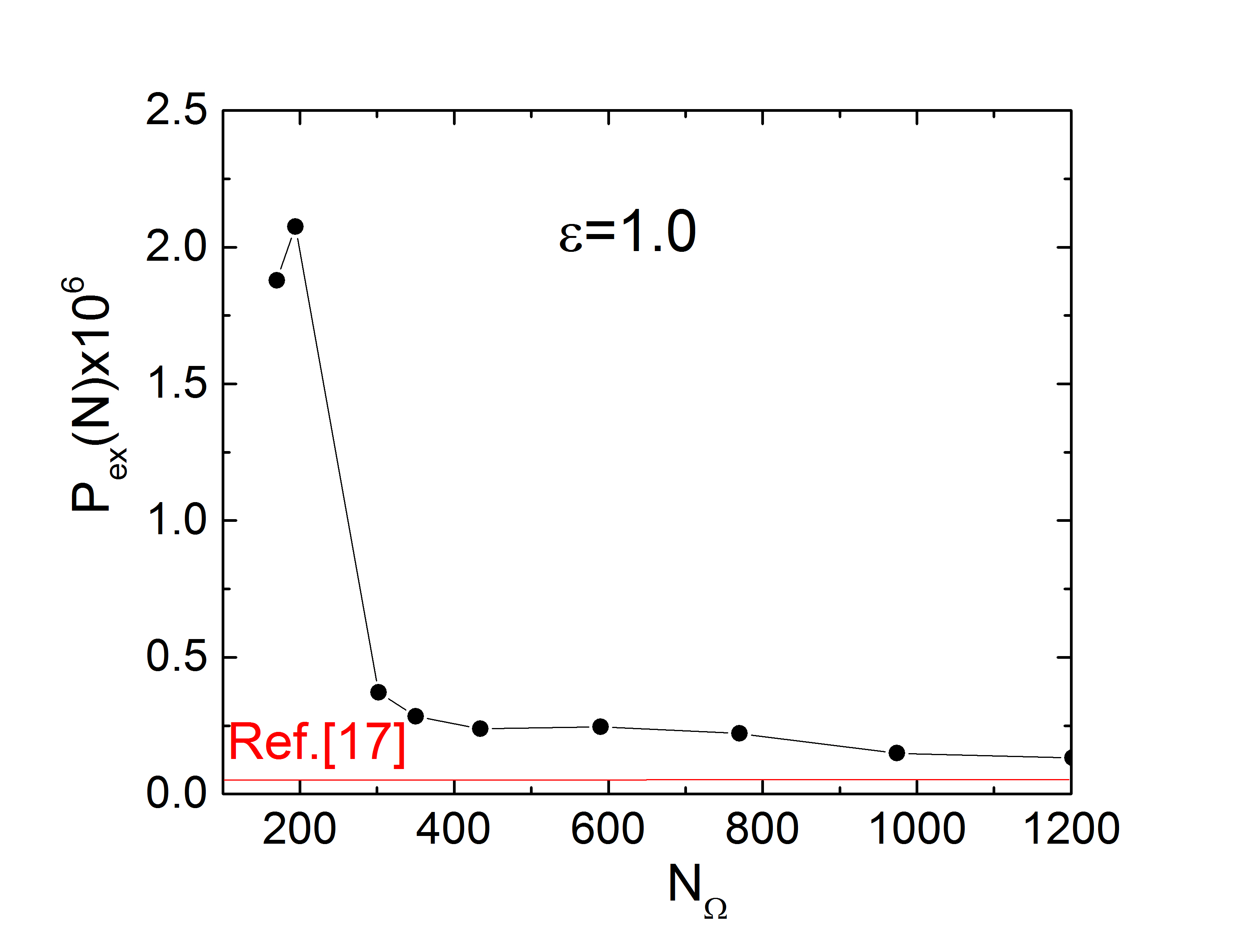} \includegraphics[width=0.4\columnwidth]{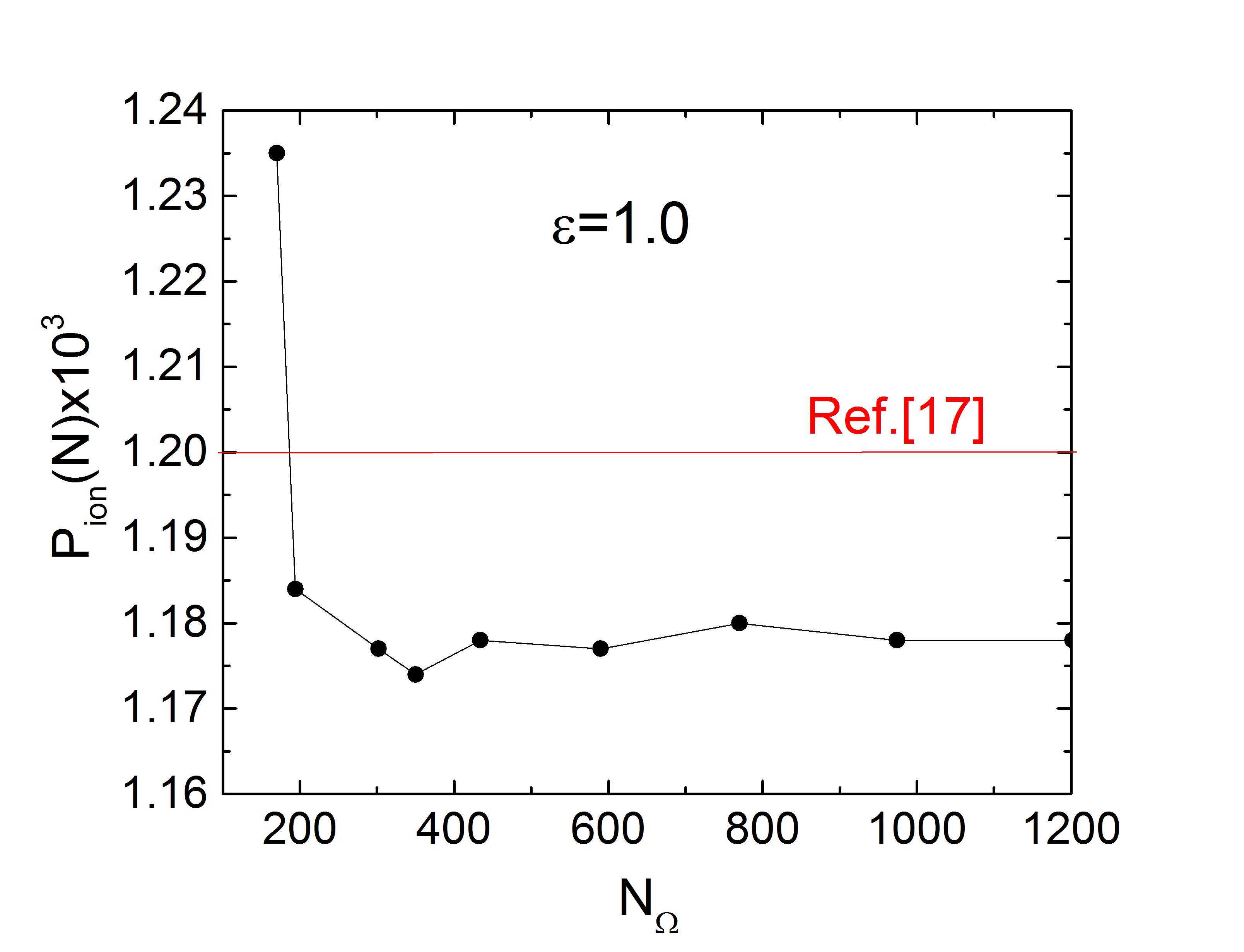}}
\end{subfigure}
\end{center}
\caption{The graphs of $P_{ex}$ and $P_{ion}$ calculated as a function of $N_\Omega$ with the 2D npDVR Lebedev scheme.}
\label{fig:Fig6}
\end{figure}

To calculate the probabilities of total excitation $P_{ex}$ and ionization $P_{ion}$ of the hydrogen atom, depending on the ellipticity of the laser field, we have used the Lebedev 2D npDVR scheme, which allowed us to carry out calculations with the largest number of basis DVR functions up to $N_\Omega = 1202$. In this case, to approximate the infinite summation (\ref{eq:discardedPn}) when calculating the probability $P_{ex}$ of total excitation (\ref{eq:PexReplaced}), we use the computational procedure (\ref{eq:PexReplaced}), (\ref{eq:discardedPn}), whose efficiency was demonstrated in Sec.\ref{Subsec:Linear} for linear polarization of the laser field. Figure \ref{fig:Fig5} illustrates this computational procedure for $\epsilon = 0.5$ and $1$. The ground state population $P_g$ and thus calculated probabilities $P_{ex}$ and $P_{ion}$ are presented in Tables \ref{tab:Table6}, \ref{tab:Table7} and \ref{tab:Table8}. Figures \ref{fig:Fig6} also depict the obtained $P_{ex}$ and $P_{ion}$ values. The data presented in Tables \ref{tab:Table6}, \ref{tab:Table7} and \ref{tab:Table8} and in Figs.\ref{fig:Fig6} convincingly demonstrate the convergence of the Lebedev 2D npDVR scheme as $N_\Omega\rightarrow +\infty$ and give the estimate $\Delta P_x \simeq \mid P_x(N_\Omega=974)-P_x(N_\Omega=1202)\mid$ of the achieved absolute accuracy in  calculating $P_g$, $P_{ex}$ and $P_{ion}$, which is improved from the values of the order $\sim 10^{-4}$ at $\epsilon = 0$ to $10^{-6}$ at $\epsilon = 1$.
It should be noted that the errors due to the finiteness of the radial box size $r_m$ and the steps of integration over the radial and time variables do not exceed this achieved accuracy. Table \ref{tab:Table9} illustrates that the radial grid with $r_m=500$ and $\Delta x=1/N_r=0.0005$ already meets this requirement when calculating $P_{ex}$ and $P_{ion}$.
The difference between the calculated values $P_g$, $P_{ex}$ and $P_{ion}$ and most accurate values \cite{Gao2019} calculated so far with alternative methods decreases with increasing ellipticity to $\epsilon=1$. However, in the entire range of $\epsilon$ variation, this difference remains an order of magnitude greater than the achieved accuracy of our calculations. We explain this deviation by the fact that we managed to significantly improve, compared to \cite{Gao2019}, the approximation of the angular part of the 3D TDSE using our 2D npDVR, which ensured fast convergence, and high computational efficiency of our algorithm where the computational time is proportional to $\sim N_r(2N_\Omega^2+200N_\Omega)$.  Perhaps, this is also due to the fact that in the pseudospectral method of \cite{Gao2019}  instead of the unperturbed bound states of the hydrogen atom $ R_{nl}(r) Y_{lm} (\theta, \phi)$ , pseudostates are used starting from $n> 16$, although there is a fairly large set of them.

\begin{table}[ht]
\caption{The probabilities $P_g$, $P_{ex}$ and $P_{ion}=1-P_g-P_{ex}$ calculated with the Lebedev 2D npDVR scheme for the laser polarization $\epsilon=0.25$. The infinite summation in $P_{ex}$ is performed according to the procedure (\ref{eq:PexReplaced}), (\ref{eq:discardedPn}). In the last row, the results of ref.\cite{Gao2019} are presented.} 
\label{tab:Table6}
\centering
\begin{indented}
\item[]\begin{tabular}{|c|c|c|c|c|c|}
\br
N & $\sum_{n=2}^{7}P_n$ &	$\sum_{n=8}^{+\infty}P_n$ &	$P_{ex}=\sum_{n=2}^{+\infty}P_n$ & $P_g$ &	$P_{ion}$\\ [0.5ex]
\hline 
170 & 4.066E-04 & 1.210E-04 & 5.276E-04 & 0.987283 & 1.219E-02\\[1ex]
194 & 2.497E-04 & 4.597E-05 & 2.956E-04 & 0.987878 & 1.183E-02\\[1ex]
302 & 3.286E-04 & 4.742E-05 & 3.760E-04 & 0.987795 & 1.183E-02\\[1ex]
350 & 2.224E-04 & 5.688E-05 & 2.792E-04 & 0.987915 & 1.181E-02\\[1ex]
434 & 4.101E-04 & 5.808E-05 & 4.682E-04 & 0.987916 & 1.162E-02\\[1ex]
590 & 4.876E-04 & 7.656E-05 & 5.642E-04 & 0.987969 & 1.147E-02\\[1ex]
770 & 5.705E-04 & 1.069E-04 & 6.774E-04 & 0.988087 & 1.124E-02\\[1ex]
974 & 6.224E-04 & 1.233E-04 & 7.457E-04 & 0.988130 & 1.112E-02\\[1ex]
1202 & 6.763E-04 & 1.341E-04 & 8.104E-04 & 0.988055 & 1.113E-02\\[1ex]
\hline\\
ref.\cite{Gao2019} & -- & -- & 1.790E-03 & 0.987480 & 1.073E-02\\[1ex]
\hline 
\end{tabular}
\end{indented}
\end{table}
\begin{table}[ht]
\caption{ The same as in table \ref{tab:Table6} for $\epsilon=0.5$.} 
\label{tab:Table7}
\centering
\begin{indented}
\item[]\begin{tabular}{|c|c|c|c|c|c|}
\br
N & $\sum_{n=2}^{7}P_n$ &	$\sum_{n=8}^{+\infty}P_n$ &	$P_{ex}=\sum_{n=2}^{+\infty}P_n$ & $P_g$ &	$P_{ion}$\\ [0.5ex]
\hline 
170 & 5.284E-05 & 5.545E-05 & 1.083E-04 & 0.994207 & 5.685E-03\\[1ex]
194 & 3.049E-05 & 1.622E-05 & 4.671E-05 & 0.993874 & 6.079E-03\\[1ex]
302 & 1.790E-05 & 7.309E-06 & 2.521E-05 & 0.994011 & 5.964E-03\\[1ex]
350 & 1.765E-05 & 1.187E-05 & 2.951E-05 & 0.993992 & 5.978E-03\\[1ex]
434 & 1.438E-05 & 3.631E-06 & 1.801E-05 & 0.993973 & 6.009E-03\\[1ex]
590 & 1.707E-05 & 4.900E-06 & 2.197E-05 & 0.993978 & 6.000E-03\\[1ex]
770 & 2.232E-05 & 9.635E-06 & 3.195E-05 & 0.993987 & 5.981E-03\\[1ex]
974 & 2.755E-05 & 1.637E-05 & 4.391E-05 & 0.993994 & 5.962E-03\\[1ex]
1202 & 3.354E-05 & 1.576E-05 & 4.930E-05 & 0.993997 & 5.954E-03\\[1ex]
\hline\\
ref.\cite{Gao2019} & -- & -- & 1.280E-04 & 0.993862 & 6.010E-03\\[1ex]
\hline 
\end{tabular}
\end{indented}
\end{table}
\begin{table}[ht]
\caption{The same as in table \ref{tab:Table6} for $\epsilon=1.0$.} 
\label{tab:Table8}
\centering
\begin{indented}
\item[]\begin{tabular}{|c|c|c|c|c|c|}
\br
N & $\sum_{n=2}^{7}P_n$ &	$\sum_{n=8}^{+\infty}P_n$ &	$P_{ex}=\sum_{n=2}^{+\infty}P_n$ & $P_g$ &	$P_{ion}$\\ [0.5ex]
\hline 
170 & 2.161E-07 & 1.663E-06 & 1.879E-06 & 0.998762826 & 1.235E-03 \\[1ex]
194 & 4.424E-07 & 1.632E-06 & 2.075E-06 & 0.998813596 & 1.184E-03 \\[1ex]
302 & 1.352E-07 & 2.361E-07 & 3.713E-07 & 0.998822676 & 1.177E-03 \\[1ex]
350 & 4.387E-08 & 2.408E-07 & 2.847E-07 & 0.998825331 & 1.174E-03 \\[1ex]
434 & 7.752E-08 & 1.607E-07 & 2.382E-07 & 0.998821665 & 1.178E-03 \\[1ex]
590 & 8.459E-08 & 1.612E-07 & 2.457E-07 & 0.998822276 & 1.177E-03 \\[1ex]
770 & 6.134E-08 & 1.600E-07 & 2.213E-07 & 0.998820162 & 1.180E-03\\[1ex]
974 & 5.407E-08 & 9.536E-08 & 1.494E-07 & 0.998821985 & 1.178E-03\\[1ex]
1202 & 3.359E-08 & 9.887E-08 & 1.325E-07 & 0.998821691 & 1.178E-03\\[1ex]
\hline\\
ref.\cite{Gao2019} & -- & -- & 2.390E-08 & 0.998799976 & 1.200E-03\\[1ex]
\hline 
\end{tabular}
\end{indented}
\end{table}
\begin{table}[ht]
\caption{The total excitation probability $P_{ex}$ and ionization yield $P_{ion}$ for the elliptically polarized laser fields with $\epsilon=0.25$ , $0.5$, and $1.0$ in terms of increasing $r_m$ and $N_r$. The calculations are performed with the Lebedev 2D npDVR scheme at $N_\Omega=770$. }
\label{tab:Table9}
\centering
\begin{indented}
\item[]\begin{tabular}{|c|c|c|c|}
\br
\multicolumn{4}{|c|}{\large{$\epsilon=0.25$}}\\ [0.5ex]
\hline
& $r_m=250$ , $N_r=1000$ & $r_m=500$ , $N_r=2000$ & $r_m=1000$ , $N_r=4000$ \\[0.5ex]
\hline 
$P_{ex}$  &  6.81E-04 &  6.77E-04 &  6.78E-04\\[1ex]
$P_{ion}$ &  1.123E-02 &  1.124E-02 &  1.124E-02\\[1ex]
\hline
\multicolumn{4}{|c|}{\large{$\epsilon=0.5$}}\\ [0.5ex]
\hline
$P_{ex}$  & 3.21E-05 &	3.20E-05 &	3.20E-05\\[1ex]
$P_{ion}$ & 5.980E-03 &	5.981E-03 &	5.982E-03\\[1ex]
\hline
\multicolumn{4}{|c|}{\large{$\epsilon=1.0$}}\\ [0.5ex]
\hline 
$P_{ex}$  &  2.40E-07 &  2.21E-07 &  2.22E-07\\[1ex]
$P_{ion}$ &  1.179E-03 &  1.180E-03 &  1.180E-03\\[1ex]
\br
\end{tabular}
\end{indented}
\end{table}

\section{Conclusion}
\label{Sec:Conclusion}

We have developed an efficient computational scheme for integrating the 3D TDSE and successfully applied it to calculate the probabilities of total excitation $P_{ex}$ and ionization $P_{ion}$ of a hydrogen atom in strong elliptically polarized laser fields. In our approach the nonseparable angular part of the 3D TDSE is  approximated with  2D npDVRs based on the 2D Lebedev and Popov cubatures for the unit sphere. Using this approach, we have calculated the probabilities $P_g$, $P_{ex}$ and $P_{ion}$ for a laser of $I=10^{14} $W/cm$^2$ and $\lambda=800$nm with absolute accuracy which is improved by increasing ellipticity of the laser field from values of the order of $\sim 10^{-4}$ for $\epsilon =0$ to $\sim 10^{-6}$ for $\epsilon=1$. We associate such improvement with a more adequate approximation of the angular part of the 3D TDSE in 2D npDVR as compared to the 2D DVR based on the direct product of 1D Gaussian DVRs where patalogic clustering of the angular grid points is present near the poles $x=y=0$. With an increase in ellipticity, our results approach the most accurate recent calculations \cite{Gao2019}; however, the difference between them remains an order of magnitude greater than the achieved accuracy of our calculations. We believe that the record accuracy in our calculations of the probabilities of excitation and ionization of a hydrogen atom in a strong elliptically polarized laser field was achieved due to the high efficiency of approximation of the angular part of the 3D TDSE in 2D npDVR, which ensured a fast convergence of the method, and its high computational efficiency with the number of simulation operations $\sim N_r (2 N_\Omega^2+200 N_\Omega) \sim N_r N_\Omega$ compared to current best practices.

We have also proposed a novel simple procedure for infinite summation of the bound state populations of the hydrogen atom in calculating the total excitation probability $P_{ex}$ and proved its accuracy by making a comparison with conventional methods.

The performed research demonstrates the potential prospects of the developed method for quantitative investigations  of the phenomena stimulated by high-intensity lasers such as HHG, ATI and photoelectron momentum distribution, where the problem, even in the case of linear polarization of the laser field, becomes essentially three-dimensional.

\ack
We are grateful to Yu. V. Popov and B. Piraux for valuable discussions and thank Xiang Gao for explaining the method and providing the data of \cite{Gao2019}. This work was supported by the Russian Science Foundation under Grant No. 20-11-20257.



\end{document}